\documentclass[useAMS,usenatbib,referee]{biom}

\usepackage{algorithm}
\usepackage{algorithmicx}
\usepackage{algpseudocode}
\usepackage{amsmath,epsfig,amssymb,bm,amsfonts}
\usepackage{xcolor,graphicx}
\usepackage{hyperref,subfigure, multirow, setspace,  booktabs}

\graphicspath{{figures/}}
\newcommand{\nvs}{\vspace{-0.26in}}

\newtheorem{Pro}{Proposition}

\def\bse{\begin{eqnarray*}}
\def\ese{\end{eqnarray*}}
\def\be{\begin{eqnarray}}
\def\ee{\end{eqnarray}}
\def\bsq{\begin{equation*}}
\def\esq{\end{equation*}}
\def\bq{\begin{equation}}
\def\eq{\end{equation}}

\def\bse{\begin{eqnarray*}}
\def\ese{\end{eqnarray*}}
\def\be{\begin{eqnarray}}
\def\ee{\end{eqnarray}}
\def\bsq{\begin{equation*}}
\def\esq{\end{equation*}}
\def\bq{\begin{equation}}
\def\eq{\end{equation}}

\def\wh{\widehat}

\def\trans{^{\rm T}}

\def\bmu{\boldsymbol\mu}

\def\beps{\boldsymbol\epsilon}

\def\bmu{\boldsymbol\mu}
\def\bg{\boldsymbol\gamma}

\def\la{\langle}
\def\btheta{\boldsymbol\theta}

\def\bbo{{\bf 0}}
\def\0{{\bf 0}}

\def\A{{\bf A}}
\def\U{{\bf U}}
\def\V{{\bf V}}

\def\B{{\bf B}}

\def\D{{\bf D}}
\def\V{{\bf V}}

\def\f{{\bf f}}
\def\h{{\bf h}}

\def\I{{\bf I}}

\def\F{{\bf F}}
\def\H{{\bf H}}
\def\m{{\bf m}}

\def\U{{\bf U}}
\def\S{{\bf S}}

\def\m{{\bf m}}

\def\X{{\bf X}}
\def\S{{\bf S}}
\def\x{{\bf x}}
\def\I{{\bf I}}

\def\y{{\bf y}}

\def\bq{\begin{equation}}
\def\eq{\end{equation}}

\def\wh{\widehat}

\def\trans{^{\rm T}}

\def\squarebox#1{\hbox to #1{\hfill\vbox to #1{\vfill}}}

\def\btheta{{\boldsymbol \theta}}

\def\bse{\begin{eqnarray*}}
\def\ese{\end{eqnarray*}}
\def\be{\begin{eqnarray}}
\def\ee{\end{eqnarray}}
\def\bsq{\begin{equation*}}
\def\esq{\end{equation*}}
\def\bq{\begin{equation}}
\def\eq{\end{equation}}
\def\wh{\widehat}

\def\trans{^\mathrm{T}}

\def\boxit#1{\vbox{\hrule\hbox{\vrule\kern6pt\vbox{\kern6pt#1\kern6pt}\kern6pt\vrule}\hrule}}

\def\red{\color{red}}

\title{High-Dimensional Multi-Study Robust Factor Model for Analyzing RNA Sequencing Data from Heterogeneous Sources}
\author{Xiaolu Jiang$^{1}$, Wei Liu$^1$\\
$^1$School of Mathematics, Sichuan University, Chengdu,  China\\{*Corresponding author.   Email: \emph{liuwei8@scu.edu.cn}}
}

\begin{document}

\label{firstpage}

\begin{abstract}
The amount of high-dimensional large-scale RNA sequencing data derived from multiple heterogeneous sources has increased exponentially in biological science. During data collection, significant technical noise or errors may occur. To robustly extract meaningful features from this type of data, we introduce a high-dimensional multi-study robust factor model, called MultiRFM, which learns latent features and accounts for the heterogeneity among sources. MultiRFM demonstrates significantly greater robustness compared to existing multi-study factor models and is capable of estimating study-specific factors that are overlooked by single-study robust factor models. Specifically, we utilize a multivariate t-distribution to model errors, capturing potential heavy tails, and incorporate both study-shared and study-specified factors to represent common and specific information among studies. For parameter estimation, we have designed a computationally efficient variational estimation approach. A step-wise singular value ratio method is proposed to determine the discrete tuning parameters. Extensive simulation studies indicate that MultiRFM surpasses state-of-the-art methods in terms of estimation accuracy across various scenarios. Real-world applications involving two RNA sequencing datasets demonstrate that MultiRFM outperforms competing methods in model fitting, prediction, and computational efficiency, significantly facilitating downstream tasks. The R package for the proposed method is publicly available at \url{https://github.com/feiyoung/MultiRFM}.
\end{abstract}

\maketitle
\section{Introduction}
In recent years, the availability of high-dimensional large-scale data from multiple sources or studies has increased exponentially in biological science. This surge has been driven by advancements in next-generation sequencing technology, which offers significant biological insights with enhanced throughput and resolution. These sources can encompass various experimental conditions, time points of sequencing samples, laboratories, and technologies. For example, \cite{xu2020differential} applied single-cell RNA sequencing technologies to measure the expression levels of tens of thousands of genes across 200,059 cells from peripheral blood mononuclear cells (PBMCs) of uninfected controls and COVID-19 patients. 
\cite{cao2019single} used sci-RNAseq3 technology to profile the transcriptomes of approximately two million cells derived from 61 embryos staged between 9.5 and 13.5 days of gestation, 
and  \cite{ben2023integration} developed a new spatial multi-modal sequencing technology, SPOTS, which achieves the measurement of RNA expression levels and protein markers within thousands of cells in each of two mouse spleen tissue sections.  Their objective was to investigate the similarities and differences in gene expression patterns and cell types in tissues from various sources.
New sequencing technologies~\citep{liu2023high} 
are continuously being developed, leading to the prevalence of multi-study, high-dimensional, and large-scale data. Given the explosion of these data types, extracting meaningful information from these valuable datasets has become an urgent issue. The challenge lies in their unique features: high noise during sequencing, erroneous values due to dropout events~\citep{li2018accurate}, significant heterogeneity among studies, high dimensionality and large sample sizes. 

Factor models inherently hold an advantage in extracting valuable information from high-dimensional, large-scale data derived from multiple studies. For instance, \cite{de2019multi} introduced the \underline{m}ulti-\underline{s}tudy \underline{f}actor \underline{a}nalysis (MSFA) model, which utilizes study-shared factors to capture common information across studies and study-specific factors to distinguish unique attributes within each study. Building upon this framework, \cite{de2021bayesian} and \cite{grabski2023bayesian} developed the Bayesian variant of MSFA, employing the MCMC algorithm for parameter estimation. However, this approach is marked by high computational costs. To enhance computational efficiency, \cite{hansen2024fast} proposed a variational estimation approach using variational approximation. \cite{argelaguet2020mofa+} extended MSFA to accommodate multi-modal single-cell data. Recently, \cite{chandra2024inferring} developed subspace factor analysis models for multi-study analysis. Nonetheless, all the aforementioned models lack robustness owing to their restrictive Gaussian assumption regarding the error term, making them unsuitable for analyzing sequencing data characterized by high noise and erroneous values.

When analyzing RNA sequencing data, the robustness of a statistical method is paramount, given the potential for significant technical noise or errors to arise at multiple stages in the sequencing pipeline. During the stage of sample preparation and handling, contamination during RNA extraction and purification can introduce noise, while RNA degradation during handling or storage can result in loss of critical information~\citep{lafzi2018tutorial}. In the library preparation stage, RNA molecules can be lost randomly, causing dropout events where expressed genes are not detected~\citep{ye2020schinter}. Additionally, the amplification step in library preparation can introduce biases, amplifying some RNA molecules more than others and thereby distorting the true gene expression levels~\citep{shi2021bias}. Furthermore, sequencing machines may introduce errors during the sequencing process~\citep{jovic2022single}. While several robust latent factor models have been developed, they primarily focus on single-study data and cannot handle the heterogeneity of data from multiple studies; see \cite{yu2019robust, he2022large, yang2023robust} and \cite{qiu2024robust}. 
Therefore, developing a  multi-study robust factor model is crucial for analyzing RNA sequencing data derived from diverse sources.

To overcome the limitations of current methods and meet practical demands, we introduce a  \underline{multi}-study \underline{r}obust  \underline{f}actor \underline{m}odel, simplified as MultiRFM. This model simultaneously tackles high-dimensional, large-scale data from multiple heterogeneous sources containing heavy-tailed noise. By utilizing a multivariate Student's t-distribution for the error term, with a degree of freedom parameter $\nu$, our model adapts dynamically to the tail behavior of errors through data-driven estimation of $\nu$. 
Second, we conduct a theoretical investigation into the model identifiability of the proposed model, which is crucial for interpretability, to uniquely determine the model parameters. Third, we propose a varitional estimation approach to estimate model parameters, implemented by a computationally efficient  
\underline{v}ariational \underline{e}xpectation-\underline{m}aximization (VEM) algorithm. Fourth, we devise a criterion based on a step-wise singular value ratio to determine the number of study-shared and study-specific factors within the MultiRFM framework. Finally, comprehensive simulation studies demonstrate that MultiRFM significantly surpasses existing methods in terms of estimation accuracy. Two real-data applications further confirm the superior model fitting and prediction accuracy, as well as the computational efficiency, of MultiRFM in analyzing noisy, high-dimensional, and large-scale RNA sequencing data obtained from various heterogeneous sources.

The subsequent sections of this paper are structured as outlined below. Section \ref{sec:model} provides an overview of the model configuration, identifiability conditions, and estimation methods. In Section \ref{sec:imp}, we present the implementation scheme of MultiRFM, including the VEM algorithm and the procedure for determining the number of factors. To assess the effectiveness of MultiRFM, Section \ref{sec:simu} conducts simulation studies, while Section \ref{sec:real} presents two real-data applications. Lastly, in Section \ref{sec:dis}, we briefly discuss potential avenues for future research in this area. Furthermore, we have seamlessly integrated MultiRFM into an efficient and user-friendly R package, which is readily available at \url{https://github.com/feiyoung/MultiRFM}.


\nvs
\section{Model and estimation}\label{sec:model}
\subsection{Proposed model}
To demonstrate the proposed model setup, we use the  single-cell RNA sequencing dataset as an example. In the single-cell RNA sequencing for tissues from  $S$ different sources, we observe a cell-by-gene matrix $\X_s \in \mathbb{R}^{n_s \times p}$ from each source $s$, where $n_s$ represents the number of cells in source $s$ and $p$ is the number of genes. We allow $p$ far exceed $n_s$. Denote the gene expression vector of cell $i$ in source $s$ as $\x_{si}$, the $i$-th column of $\X_s^{\trans}$. To address potential heavy tails in the data, we propose a multi-study robust  latent factor model, formulated as follows:
\begin{equation}\label{eq:model1}
\x_{si} = \bmu_s + \A\f_{si} + \B_s \h_{si} + \beps_{si},
\end{equation}
where the error term $\beps_{si}$ is independently and identically distributed (i.i.d.) according to a multivariate Student's t-distribution, denoted as $MVT_{p}(\nu,\mathbf{0}, \Lambda_s)$, where $\Lambda_s=\mathrm{diag}(\lambda_{s1}, \cdots, $ $\lambda_{sp})$. Here, $\nu$ represents the degree of freedom, $\mathbf{0}$ is the center parameter, and $\Lambda_s$ is the scale parameter. $\bmu_s$ is the shift mean, capturing the mean expression levels of all cells in source $s$. The matrix $\A \in \mathbb{R}^{p \times q}$ captures the study-shared loading effects, while $\B_s \in \mathbb{R}^{p \times q_s}$ captures the study-specific loading effects. The latent factors $\f_{si} \stackrel{i.i.d.}\sim N(\mathbf{0}, \I_q)$ and $\h_{si} \stackrel{i.i.d.}\sim N(\mathbf{0}, \I_{q_s})$ are  the study-shared and study-specific factors, respectively, where $\I_q$ is an identity matrix of dimension $q$. We assume $(\f_{si}, \h_{si}, \varepsilon_{si})$ are mutually independent. Unlike existing multi-study latent factor models \citep{de2019multi,grabski2023bayesian,hansen2024fast}, our model provides a robust framework for analyzing multi-study data with potential  heavy-tailed distributions. 
\nvs
\subsection{Model identifiability}
The issue of identifiability frequently poses a challenge in latent factor models~\citep{de2019multi}. Consequently, our initial step involves exploring identifiable conditions to ensure a meaningful interpretation of model \eqref{eq:model1}.
To achieve uniqueness of model parameters, we impose a comprehensive set of identifiability conditions. Let $q_{\min}=\min_s q_s$.
The conditions are as follows:
\begin{itemize}
  \item[] (A1) {\it Diagonal structure and positivity: $(\A,\B_1)^{\trans}(\A, \B_1)$ and $\{\B_s^{\trans}\B_s, 2\leq s\leq S\}$ are diagonal matrices with decreasing positive diagonal entries. Furthermore, the first nonzero element in each column of $\A$ and $\B_s$ (for $s \leq S$) is positive.} 
  \item[](A2) {\it Dimensionality constraint: $p-1 > q + \sum_{s=1}^S q_s$.} 
  \item[](A3) {\it Study-specific factors: For any $k \leq q_{\min}$, there exist studies $s_1$ and $s_2$ such that the $k$-th column of $\B_{s_1}$ differs from that of $\B_{s_2}$.} 
  \item[](A4) {\it Constraint of the degree of freedom: $\nu>2$.} 
\end{itemize}
Condition (A1) prevents rotational invariance of $\A$ and $\B_s$, also used in \cite{bai2013principal} and \cite{GFMLiu}. Condition (A2)  ensures that the total number of latent factors  is less than the number of observed variables minus one, a condition typically met in practice. Condition (A3) guarantees that $h_{sik}$ truly represents a study-specific factor, preventing it from becoming a study-shared factor if $\B_{s, .k}$s were identical across all studies. Condition (A4) ensures the existence of error's covariance matrix due to $\mathrm{cov}(\beps_{si})=\frac{\nu}{\nu-2}\Lambda_s$.

By adhering to above conditions, we state the identifiability in the following proposition, with its proof deferred to Supplementary Materials.
\nvs
\begin{Pro}
 Under Conditions (A1)--(A4), model \eqref{eq:model1} is identifiable.
\end{Pro}
\nvs
\subsection{Estimation}
Let $\X = (\X_s, s=1,\cdots, S), \F_s = (\f_{s1}, \cdots, \f_{sn_s})^{\trans},\H_s = (\h_{s1}, \cdots, \h_{sn_s})^{\trans}, \F = (\F_s, s=1, \cdots, S)$ and $\H=(\H_s, s=1, \cdots, S)$, and denote $\btheta=(\A, \B_s, \bmu_s, \Lambda_s, \nu, s=1, \cdots, S)$, the collection of all model parameters. Under the assumptions in model \eqref{eq:model1}, the full likelihood is written as
\begin{eqnarray}
 &&   L(\btheta; \X, \F, \H) = \Pi_{s=1}^S\Pi_{i=1}^{n_s} P(\x_{si}|\f_{si}, \h_{si}) P(\f_{si}) P(\h_{si}) \nonumber \\
  &=& \Pi_{s=1}^S\Pi_{i=1}^{n_s} [1+\frac{1}{\nu}(\x_{si}-\bmu_s - \A\f_{si} - \B_s \h_{si} )^{\trans}\Lambda_s^{-1}(\x_{si}-\bmu_s - \A\f_{si} - \B_s \h_{si}) ]^{-(\nu+p)/2} \nonumber \\
  & & \times \Pi_{s=1}^S\Pi_{i=1}^{n_s}\left\{ C_p(\nu) |\Lambda_s|^{-1/2}\exp(-\frac{1}{2}\f_{si}^{\trans}\f_{si}) \exp(-\frac{1}{2}\h_{si}^{\trans}\h_{si})\right\} \times c, \label{eq:fulllike}
\end{eqnarray}
 where $C_p(\nu)=(\pi \nu)^{-p/2} \Gamma\{(\nu+p)/2\}/\Gamma(\nu/2)$ with $\Gamma(\cdot)$ being the Gamma function and $c$ is a constant independent of model parameters. Theoretically, by integrating the latent random matrices $\F$ and $\H$ in \eqref{eq:fulllike}, we are able to obtain the maximum likelihood estimator by maximizing the observed log-likelihood function
$l_o(\btheta;\X)=\ln \int L(\btheta; \X, \F, \H)d\F d\H.$
Due to the high nonlinearity caused by the multivariate t-distribution and the complexity of high-dimensional integration, the optimization of $l_o(\btheta; \X)$ becomes intractable. Therefore, we aim to construct a tractable surrogate function of $l_o(\btheta;\X)$ by leveraging variational approximation techniques~\citep{blei2017variational, liu2024high} and the property of the convex function.

First, we derive the full log-likelihood, given by
\begin{eqnarray*}
&& l(\btheta; \X, \F, \H)  \\
  &=& -\sum_{s,i}\frac{\nu+p}{2} \ln\left[1+\frac{1}{\nu} (\x_{si}-\bmu_s - \A\f_{si} - \B_s \h_{si} )^{\trans}\Lambda_s^{-1}(\x_{si}-\bmu_s - \A\f_{si} - \B_s \h_{si}) \right] \\
  &+& \sum_{s} \left\{n_s\ln C_{p}(\nu)  - \frac{n_s}{2} \ln|\Lambda_s| \right\}-\sum_{s,i} \frac{1}{2}(\f_{si}^{\trans}\f_{si} + \h_{si}^{\trans}\h_{si})+c.
\end{eqnarray*}
Let $q(\F,\H;\bg)$ be a variational density function for $(\F,\H)$ with variational parameter $\bg$. By the variational inference theory~\citep{blei2017variational}, we know the evidence lower bound $l_{elbo}(\btheta, \bg)=\mathrm{E}_q l(\btheta; \X, \F,\H)- \mathrm{E}_q \ln q(\F,\H;\bg)\leq l_o(\btheta;\X)$, with equality holding if and only if $q(\F,\H;\bg)$ is the posterior distribution of $(\F,\H)$, where $\mathrm{E}_q$ represents the expectation with respect to $(\F, \H)$ based on the density function $q(\F,\H;\bg)$. The variational inference usually involves calculating the explicit form of $l_{elbo}(\btheta, \bg)$ and then obtain the maximizer of $\btheta$. However, in our model, the computation of $\mathrm{E}_q l(\btheta; \X, \F,\H)$ is very challenging due to the term $\ln (1+g(\f_{si}, \h_{si}))$, where $g(\f_{si}, \h_{si}) = \frac{1}{\nu} (\x_{si}-\bmu_s - \A\f_{si} - \B_s \h_{si} )^{\trans}\Lambda_s^{-1}(\x_{si}-\bmu_s - \A\f_{si} - \B_s \h_{si})$. 

Fortunately, $-\ln(1+x)$ is a convex function with respect to $x$, thus, we have $\mathrm{E}_q (-\ln (1+g(\f_{si}, \h_{si}))) \geq -\ln (1+ \mathrm{E}_qg(\f_{si}, \h_{si})))$ by applying Jensen's inequality. Therefore, we obtain $l_{elbo}(\btheta, \bg) \geq l(\btheta, \bg)$, where
\begin{eqnarray*}
&&  l(\btheta, \bg)\\
  &= &-\sum_{s,i}\frac{\nu+p}{2} \ln\left[1+\frac{1}{\nu}\mathrm{E}_q  (\x_{si}-\bmu_s - \A\f_{si} - \B_s \h_{si} )^{\trans}\Lambda_s^{-1}(\x_{si}-\bmu_s - \A\f_{si} - \B_s \h_{si}) \right] \\
  &+& \sum_{s} \left\{n_s\ln C_{p}(\nu)  - \frac{n_s}{2} \ln|\Lambda_s| \right\}-\sum_{s,i} \frac{1}{2}\mathrm{E}_q (\f_{si}^{\trans}\f_{si} + \h_{si}^{\trans}\h_{si})- \mathrm{E}_q \ln q(\F, \H;\bg)+c.
\end{eqnarray*}
The mean field variational family is the most commonly utilized in variational inference, as noted in \citep{blei2017variational}. In this context, we also adopt $q(\F,\H;\bg)$ as a mean field variational family, which is defined as 
$q(\F,\H;\bg) = \Pi_{s=1}^S\Pi_{i=1}^{n_s} N(\f_{si}; \m_{f,si},\S_{f,si}) N(\h_{si}; \m_{h,si},\S_{h,si})$, where  $\bg=(\m_{f,si},\S_{f,si},$ $ \m_{h,si},\S_{h,si}, s=1,\cdots, S, i=1,\cdots, n_s)$ and  $N(\f; \m,\S)$ represents the multivariate normal density function of $\f$ with mean vector $\m$ and covariance matrix $\S$.
By taking a simple expectation, we can derive the explicit form of $l(\btheta, \bg)$, as follows:
\begin{eqnarray}
&& \hspace{-1em} l(\btheta, \bg) \label{eq:lbtbg} \\
  &&\hspace{-2em}= -\sum_{s,i}\frac{\nu+p}{2} \ln\bigg[1+\frac{1}{\nu}  (\x_{si}-\bmu_s - \A\m_{f,si} - \B_s \m_{h,si} )^{\trans}\Lambda_s^{-1}(\x_{si}-\bmu_s - \A\m_{f,si} - \B_s \m_{h,si}) \nonumber \\
  &&\hspace{-2em} +\frac{1}{\nu}\left\{ \mathrm{Tr}(\A^{\trans}\Lambda^{-1}_s\A \S_{f,si}) +\mathrm{Tr}(\B_s^{\trans}\Lambda^{-1}_s\B_s \S_{h,si}) \right\}\bigg] + \sum_{s} \left\{n_s\ln C_{p}(\nu)  - \frac{n_s}{2} \ln|\Lambda_s| \right\} \nonumber\\
  &&\hspace{-2em} - \sum_{s,i} \frac{1}{2} \left\{\mathrm{Tr}(\m_{f,si}\m_{f,si}^{\trans} + \S_{f,si}) + \mathrm{Tr}(\m_{h,si}\m_{h,si}^{\trans} + \S_{h,si})\right\}+\frac{1}{2}\sum_{s,i} (\ln|\S_{f,si}|+\ln|\S_{h,si}|)+c. \nonumber
\end{eqnarray}
We refer to $l(\btheta, \bg)$ as the variational lower bound of $l_o(\btheta;\X)$. This variational lower bound function is more straightforward to optimize compared to the observed log-likelihood function $l_o(\btheta;\X)$. Therefore, we derive the estimator of $\btheta$ as $\wh\btheta$, where $\wh\btheta$ is obtained by maximizing the function $ l(\btheta, \bg)$ over the parameters $(\btheta,\bg)$.

\section{Implementation} \label{sec:imp}
To implement MultiRFM, we devise a variational EM algorithm on the basis of the variational lower bound function $l(\btheta, \bg)$ and introduce a
criterion for determining the number of factors.

With the variational lower bound $l(\btheta, \bg)$ available, the algorithm is straightforwardly designed.
Let $\phi_{si}(\btheta, \bg)= 1+\frac{1}{\nu}  (\x_{si}-\bmu_s - \A\m_{f,si} - \B_s \m_{h,si} )^{\trans}\Lambda_s^{-1}(\x_{si}-\bmu_s - \A\m_{f,si} - \B_s \m_{h,si})+\frac{1}{\nu}\left\{ \mathrm{Tr}(\A^{\trans}\Lambda^{-1}_s\A \S_{f,si}) +\mathrm{Tr}(\B_s^{\trans}\Lambda^{-1}_s\B_s \S_{h,si}) \right\}$. In the variational E-step, taking derivatives of $l(\btheta, \bg)$ with respect to the variational parameters and setting them to zeros, we obtain
\begin{eqnarray}
   \S_{f,si} &=& \left( \frac{\nu+p}{\nu\phi_{si}(\btheta, \bg)} \A^{\trans} \Lambda_s^{-1} \A + \I_q\right)^{-1} \label{eq:Sf} \\
  \m_{f,si} &=& \S_{f,si}\frac{\nu+p}{\nu\phi_{si}(\btheta, \bg)} \A^{\trans} \Lambda_s^{-1} (\x_{si}-\bmu_s -  \B_s \m_{h,si}) \\
    \S_{h,si} &=& \left( \frac{\nu+p}{\nu\phi_{si}(\btheta, \bg)} \B_s^{\trans} \Lambda_s^{-1} \B_s + \I_{q_s}\right)^{-1} \\
    \m_{h,si} &=& \S_{h,si}\frac{\nu+p}{\nu\phi_{si}(\btheta, \bg)} \B_s^{\trans} \Lambda_s^{-1} (\x_{si}-\bmu_s -  \A \m_{f,si}). \label{eq:mh}
\end{eqnarray}

Similarly, in the variaitonal M-step, taking derivatives of $l(\btheta, \bg)$ with respect to the model parameters and setting them to zeros, we obtain
\begin{eqnarray}
  \bmu_s &=& \frac{1}{\sum_{i=1}^{n_s}\frac{1}{\phi_{si}(\btheta, \bg)}}\sum_{i=1}^{n_s} \frac{1}{\phi_{si}(\btheta, \bg)} (\x_{si}- \A\m_{f,si} - \B_s \m_{h,si} ), \label{eq:bmu} \\
  \mathrm{vec}(\A) &=&  {\left[ \sum_{s,i} \frac{[ (\I_q \otimes \Lambda_s^{-1})\{ (\m_{f,si}\m_{f,si}^{\trans} + \S_{f,si})\otimes \I_p\}] }{\phi_{si}(\btheta, \bg)} \right]^{-1}} \sum_{s,i} \frac{\mathrm{vec}(\Lambda_s^{-1}\tilde\x_{si} \m_{f,si}^{\trans})}{\phi_{si}(\btheta, \bg)}, \label{eq:A}\\
   \B_s &=&  \sum_i\frac{1}{\phi_{si}(\btheta, \bg)}\breve\x_{si} \m_{h,si}^{\trans}\left\{ \sum_{i} \frac{1}{\phi_{si}(\btheta, \bg)} (\m_{si}\m_{si}^{\trans} + \S_{h,si})\right\}^{-1},  \\
   \mathrm{diag}(\Lambda_s) &=& \frac{1}{n_s} \sum_i \frac{\nu+p}{\nu\phi_{si}(\btheta, \bg)} \left\{ \mathrm{diag}(\y_{si}\y_{si}^{\trans} + \A\S_{f,si}\A^{\trans} + \B_s\S_{h,si}\B_s^{\trans}) \right\},  \label{eq:Lambdas}\\
   \nu &=& \arg\max_{\nu}  l(\btheta, \bg), \label{eq:tau}
\end{eqnarray}
where $\tilde\x_{si} = \x_{si}-\bmu_s - \B_s \m_{h,si}$, $\breve\x_{si}=\x_{si}-\bmu_s - \A\m_{f,si}$, $\y_{si} = \x_{si}-\bmu_s - \A\m_{f,si} - \B_s \m_{h,si} $, $\mathrm{vec}(\bullet)$ represents an operator that converts a  matrix  into a column vector by vertically stacking its columns, and $\mathrm{diag}(\bullet)$ is the operator that extracts the diagonal elements of a square matrix  to a column vector. Despite the right-hand side of Equation \eqref{eq:A} containing the inverse of a $(pq) \times (pq)$ matrix, this matrix is structured as a block diagonal matrix comprising $p$  submatrix blocks of dimension $q$. Consequently, the computational complexity of the inverse is reduced to $O(pq^3)$ rather than $O(p^3q^3)$. To speed up the computation, we use a grid search strategy to obtain the solution for $\nu$.

Through the iterative expressions, we can intuitively understand the robustness of this method. 
Notably, the discrepancy between each observed vector $\x_{si}$ and $\bmu_s + \A\f_{si}+\B_s\h_{si}$  in relation to  the expectation of $(\f_{si},\h_{si})$ is quantified by $\phi_{si}(\btheta, \bg)= 1+\frac{1}{\nu}\mathrm{E}_q  (\x_{si}-\bmu_s - \A\f_{si} - \B_s \h_{si} )^{\trans}\Lambda_s^{-1}(\x_{si}-\bmu_s - \A\f_{si} - \B_s \h_{si})$. When $\x_{si}$  deviates significantly from $\bmu_s + \A\f_{si}+\B_s\h_{si}$, $\phi_{si}(\btheta, \bg)$ takes a large value. 
Thus, the inclusion of the term $\frac{1}{\phi_{si}(\btheta, \bg)}$ in Equations \eqref{eq:bmu}--\eqref{eq:Lambdas} down-weights observations that significantly deviate from the expected pattern, thereby reducing their impact and enhancing the overall stability of the method in estimating the model parameters.

While $\phi_{si}(\btheta, \bg)$ enhances the robustness of our method, it introduces a dependency of the right-hand side of Equations \eqref{eq:Sf}--\eqref{eq:Lambdas} on the parameters $\btheta$ and $\bg$ via the function $\phi_{si}(\btheta, \bg)$. To facilitate iterative updates of the parameters while maintaining this robustness, we adopt a fixed-point iteration approach. Specifically, we fix $\phi_{si}(\btheta, \bg)$ at its value from the previous iteration, denoted as $\phi_{si}(\btheta^{(t-1)}, \bg^{(t-1)})$, where $^{(t-1)}$ indicates the $(t-1)$-th iteration. 
We encapsulate the entire process in Algorithm S1 of Supplementary Materials, which possesses  linear computational complexity with respect to both $n$ and $p$. Upon convergence of the algorithm, we utilize the variational posterior expectation $\wh\m_{f,si}$ and $\wh\m_{h,si}$ as the estimators for $\f_{si}$ and $\h_{si}$, respectively. Consequently, $\wh\F_s=(\wh\m_{f,s1}, \cdots, \wh\m_{f,sn_s})^{\trans}$ and $\wh\H_s=(\wh\m_{h,s1}, \cdots, \wh\m_{h,sn_s})^{\trans}$.
\nvs
\subsection{Determining the number of latent factors}\label{sec:select}
The successful implementation of MultiRFM requires to select the number of factors that are shared among studies ($q$) and those that are specific to individual studies ($q_s$). In the realm of multi-study data, the study-shared factors, which integrate information across all data sources, typically manifest a stronger signal than study-specific factors. To address this distinctive characteristic, we introduce a \underline{s}tep-wise \underline{s}ingular \underline{v}alue \underline{r}atio (SSVR) approach. Firstly, we adapt the single-study SVR method presented in \cite{liu2024high} to a multi-study context, utilizing this refined criterion to ascertain the optimal value of $q$. Following the determination of $q$, we employ an analogous criterion to identify $q_s$. Specifically,  we fit MultiRFM  with upper bounds on the number of factors, i.e., $q = q_{\max}$ and $q_s = q_{s,\max}$. Denote the estimator of the study-shared loading matrix as $\widehat{\mathbf{A}}^{(\max)}$.
Next, we estimate the number of study-shared factors by $\wh q=\arg\max_{k \leq q_{\max}-1} \frac{\xi_k(\wh\A^{(\max)})}{\xi_{k+1}(\wh\A^{(\max)})}$, where $\xi_k(\wh\A^{(\max)})$ is the $k$-th largest singular value of $\wh\A^{(\max)}$.  Subsequently, we refit our model with  the selected $\wh q$ and upper bounds on the number of study-specific factors $q_s=q_{s,\max}$. Subsequently, we compute the singular value ratio for the estimated study-specific loading matrices $\widehat{\mathbf{B}}_{s}^{(\max)}$ to determine $\widehat{q}_s$, applying the same SVR principle. The results in simulation study show that the proposed criterion works well.

\nvs
\section{Simulation study}\label{sec:simu}
\subsection{Compared methods and metrics}
In this section, we demonstrate the effectiveness of our proposed MultiRFM through extensive simulation studies. To emphasize its advantages, we compare MultiRFM with various other art-of-state methods in the literature, including 
\begin{itemize}
    \item[](1)  multi-study factor analysis model (MSFA)~\citep{de2019multi};
    \item[](2) Bayesian multi-study factor analysis model (BMSFA)~\citep{de2021bayesian};
    \item[](3) variational inference for  multi-study factor analysis model, owing two versions utilizing different algorithms, denoted as MSFA-CAVI and MSFA-SVI, respectively~\citep{hansen2024fast};
    \item[](4) robust two-stage estimation (RTS)~\citep{he2022large}.
\end{itemize} 
Among the aforementioned methods, both MSFA and BMSFA are implemented within the R package {\bf MSFA} (version 0.86). MSFA-CAVI and MSFA-SVI are implemented in the R package {\bf VIMSFA} (version 0.1.0). In contrast, the implementation of RTS is available in the supplementary files of the original publication~\citep{he2022large}. Notably, RTS is a robust factor model specifically designed for single-study data, meaning it disregards study-specific factors and solely estimates loadings and factors that are shared across studies.

To assess the similarity between two matrices $\wh\D$ and $\D$, we employ the trace statistic, which is defined as $\mathrm{Tr}(\wh\D,\D) = \frac{\mathrm{Tr}\{\D^{\trans} \wh\D(\wh\D^{\trans}\wh\D)^{-1}\wh\D^{\trans}\D\}}{\mathrm{Tr}(\D^{\trans}\D)}$ ~\citep{nie2024high}. This metric ranges from 0 to 1, with higher values indicating greater similarity.
 We gauge the accuracy of the estimated $\wh\A$  using the trace statistic $\mathrm{Tr}(\wh\A,\A_0)$, denoted by $\mathrm{Tr}_{\A}$. For each study-specific loading matrix $\wh\B_s$, we compute its trace statistic with the corresponding true matrix $\B_{s0}$, and then average these statistics across all studies, i.e., $\mathrm{MTr}_{\B}=\frac{1}{S}\sum_{s=1}^{S}\mathrm{Tr}(\wh\B_s,\B_{s0})$, to obtain an overall measure of accuracy. We also use the mean trace statistics, $\mathrm{MTr}_{\F}$ and $\mathrm{MTr}_{\H}$, to evaluate the estimation performance of the factor matrices $\{\F_s\}_{s=1}^S$ and $\{\H_s\}_{s=1}^S$, respectively.

\subsection{Simulation settings and results}
We assess the performance of MultiRFM and other comparative methods across five scenarios, examining various facets of their effectiveness. In all scenarios, we set {\red $S=2, (n_1, n_2)=(150,200), p=500$}  $q=3$ and $q_s=2$ unless otherwise specified, and repeat the evaluation process 100 times to ensure the  reliability.

\noindent\underline{Scenario 1}.
In this scenario, we examine the influence of the tail behavior of errors drawn from an i.i.d. multivariate t-distribution. The objective of this analysis is to comprehend how the heavy-tailed nature of the t-distribution affects MultiRFM and the methods being compared. We simulate data from model \eqref{eq:model1}. Specifically, we generate $\bmu_{s0} \stackrel{i.i.d.}\sim N(\bbo, \I_p), \f_{si} \stackrel{i.i.d.}\sim  N(\bbo, \I_q)$ and $\h_{si} \stackrel{i.i.d.}\sim  N(\bbo, \I_{q_s})$. To generate $\A_0$ and $\B_{10}$, we first generate $\breve{\B}_1=(\breve{b}_{1jk})\in \mathbb{R}^{p\times (q+q_1)}$ with $\breve b_{1jk} \stackrel{i.i.d.}\sim N(0,1)$, then perform SVD, i.e., $\breve\B_1=\U_1 \L_1 \V_1^{\trans}$ with the first nonzero entry of each column of $\U_1$ positive, and let $\bar\B_1= \rho_A \U_1 \L_1$. Let $\A_0$ be the first $q$ columns of $\bar\B_1$ and $\B_{10}$ be the last $q_1$ columns of $\bar\B_1$. For $1<s \leq S$, we generate matrix $\breve{\B}_s\in \mathbb{R}^{p\times q_s}$ with SVD $\breve{\B}_s=\U_s \L_s \V_s^{\trans}$ by the same way, and let $\B_{s0}=\rho_B \U_s \L_s$. $\rho_A$ and $\rho_B$ are designed to control the signal strength of study-shared and study-specified factors, respectively. In this scenario, we fix $(\rho_A,\rho_B)=(5,5)$. Note that $\A_0$ and $\{\B_{s0}\}_{s=1}^S$ satisfy the identifiable conditions (A1)--(A3) given in Section \ref{sec:model}. 
After generating $\bmu_{s0}, \A_0$ and $\B_{s0}$, they are fixed in  repetition. 
Finally, we generate $\beps_{si} \stackrel{i.i.d.}\sim MVT_p(\nu, \bbo, \I_p)$ by setting $\nu=2,3$ or $20$, each representing distinct characteristics of the tail distribution. A lower value of $\nu$ (e.g., $2$ or $3$) implies a heavier tail, whereas a higher value (e.g., $20$) corresponds to a lighter tail.

The comparison of estimation performance for loading matrices and factor matrices is presented in Table \ref{tab:sce1}, yielding the following key insights. Firstly, by comparing MultiRFM with existing methods designed for multiple studies, we observe that MultiRFM demonstrates a significant advantage over these methods when dealing with heavy-tailed errors ($\nu=2$ or 3). Specifically, MultiRFM achieves the highest trace statistics, surpassing other methods in estimating both loading matrices and factor matrices. Notably, MultiRFM's superiority is even more pronounced in the extreme heavy-tailed scenario with $\nu=2$, where all other non-robust methods fail to produce satisfactory estimates for the study-specific loadings and factors. The competing methods perform well for the study-shared factor part but poorly for the study-specified factor part, as the data offers more information for the study-shared part compared to the study-specified factor part. Conversely, in the lighter tail scenario ($\nu=20$), all methods perform more closely and consistently well in estimating factor matrices, while MultiRFM still outperforms the compared methods in estimating loading matrices, demonstrating its robust performance across various conditions. Secondly, by comparing MultiRFM with RTS, which is designed for robust single-study factor analysis, our observations indicate that RTS demonstrates robustness to the various tail settings and performs comparably to MultiRFM in estimating the study-shared loading and factor matrix across all considered cases. However, a notable limitation of RTS is its inability to provide the estimates for study-specified factor and loading matrices $\H_s$ and $\B_s$, overlooking the study-specific information contained within the data.

\begin{table}[htbp]  
    \centering\small  
    \caption{Comparison of MultiRFM and other competing methods for parameter estimation  in Scenario 1. Reported are average (standard deviation) of (mean) trace statistics. Note that RTS is unable to estimate the study-specified factor ($\H_s$) and loading ($\B_s$) matrices.}\label{tab:sce1} 
    \begin{tabular}{l l cccccc}
        \toprule  
        \multicolumn{1}{c}{$\nu$} & {Metric} & {MultiRFM} & {MSFA} & {BMSFA} & {MSFA-CAVI} & {MSFA-SVI} & {RTS} \\  
        \cmidrule(lr){1-8}  
        \multicolumn{1}{c}{$\nu=2$} & $\mathrm{Tr}_{\A}$ & 0.990 & 0.913 & 0.921 & 0.883 & 0.884 & 0.997 \\  
        & & {(3.8e-2)} & {(7.0e-2)} & {(4.7e-2)} & {(7.6e-2)} & {(7.5e-2)} & {(7.9e-4)} \\  
        & $\mathrm{MTr}_{\F}$ & 0.963 & 0.964 & 0.968 & 0.903 & 0.904 & 0.970 \\  
        & & {(6.1e-2)} & {(4.3e-2)} & {(2.4e-2)} & {(1.1e-1)} & {(1.1e-1)} & {(1.2e-2)} \\  
        & $\mathrm{MTr}_{\B}$ & 0.956 & 0.525 & 0.558 & 0.508 & 0.547 & {-} \\  
        & & {(4.9e-2)} & {(1.8e-1)} & {(1.8e-1)} & {(2.3e-1)} & {(2.0e-1)} & {(-)} \\  
        & $\mathrm{MTr}_{\H}$ & 0.890 & 0.375 & 0.380 & 0.354 & 0.378 & {-} \\  
        & & {(1.0e-1)} & {(1.3e-1)} & {(1.3e-1)} & {(1.6e-1)} & {(1.4e-1)} & {(-)} \\  
        \midrule  
        \multicolumn{1}{c}{$\nu=3$} & $\mathrm{Tr}_{\A}$ & 0.999 & 0.906 & 0.907 & 0.906 & 0.906 & 0.997 \\  
        & & {(9.0e-5)} & {(4.6e-3)} & {(4.7e-3)} & {(4.6e-3)} & {(4.8e-3)} & {(6.7e-4)} \\  
        & $\mathrm{MTr}_{\F}$ & 0.985 & 0.985 & 0.985 & 0.983 & 0.983 & 0.984 \\  
        & & {(3.3e-3)} & {(2.4e-3)} & {(2.4e-3)} & {(3.8e-3)} & {(3.9e-3)} & {(3.4e-3)} \\  
        & $\mathrm{MTr}_{\B}$ & 0.973 & 0.762 & 0.793 & 0.796 & 0.805 & {-} \\  
        & & {(2.7e-3)} & {(6.1e-2)} & {(5.4e-2)} & {(5.5e-2)} & {(5.5e-2)} & {(-)} \\  
        & $\mathrm{MTr}_{\H}$ & 0.940 & 0.695 & 0.694 & 0.720 & 0.745 & {-} \\  
        & & {(8.1e-3)} & {(1.3e-1)} & {(1.3e-1)} & {(1.3e-1)} & {(1.4e-1)} & {(-)} \\  
        \midrule  
        \multicolumn{1}{c}{$\nu=20$} & $\mathrm{Tr}_{\A}$ & 0.999 & 0.887 & 0.887 & 0.887 & 0.887 & 0.998 \\  
        & & {(9.7e-5)} & {(3.0e-3)} & {(3.0e-3)} & {(3.0e-3)} & {(3.1e-3)} & {(7.4e-4)} \\  
        & $\mathrm{MTr}_{\F}$ & 0.989 & 0.990 & 0.990 & 0.988 & 0.987 & 0.989 \\  
        & & {(3.4e-3)} & {(6.3e-4)} & {(6.2e-4)} & {(3.5e-3)} & {(3.5e-3)} & {(3.4e-3)} \\  
        & $\mathrm{MTr}_{\B}$ & 0.974 & 0.828 & 0.861 & 0.868 & 0.862 & {-} \\  
        & & {(2.3e-3)} & {(2.6e-2)} & {(2.3e-2)} & {(6.1e-3)} & {(7.2e-3)} & {(-)} \\  
        & $\mathrm{MTr}_{\H}$ & 0.966 & 0.911 & 0.911 & 0.962 & 0.954 & {-} \\  
        & & {(5.4e-3)} & {(8.8e-3)} & {(9.3e-3)} & {(5.2e-3)} & {(7.9e-3)} & {(-)} \\  
        \bottomrule  
    \end{tabular}  
    \label{tab:data} 
\end{table}

\noindent\underline{Scenario 2}. In addition to the multivariate t-distribution, we further explore other distributions of the error term to assess the robustness of MultiRFM against other methods. We examine three scenarios, each corresponding to a distinct error distribution.
In the first scenario, we generate $\beps_{si}$ i.i.d. from a multivariate normal distribution, specifically, $\varepsilon_{si} \sim MVN(\mathbf{0}, \mathbf{I}_p)$. In the second scenario, each entry of $\beps_{si}$ is generated i.i.d. from a centered exponential distribution, $\varepsilon_{sij} \sim \mathrm{Exp}(1) - 1$, which is skewed. In the third scenario, each entry $\varepsilon_{sij}$ of $\beps_{si}$ is generated i.i.d. from a centered Pareto distribution. This involves first generating $z_j$ from a Pareto distribution with density function $f(z) = \frac{\alpha}{z^{\alpha+1}}$ and $\alpha = 2$, and then setting $\varepsilon_{sij} = z_j - \frac{\alpha}{1-\alpha}$. Notably, when $\alpha = 2$, $\varepsilon_{sij}$ exhibits heavy tails and its second-order moment does not exist. The other settings remain consistent with those in Scenario 1.

As indicated in Table \ref{tab:diffdistri}, MultiRFM exhibits robustness against different distributions of the error term. 
Specifically, across the cases considered, MultiRFM outperforms other methods  designed for multi-study factor analysis by a wide margin in terms of the study-shared loading matrix ($\mathrm{Tr}_{\A}$), study-specific loading matrix ($\mathrm{MTr}_{\B}$), and study-specific latent factor matrix ($\mathrm{MTr}_{\H}$). Notably, for the heavy-tailed Pareto distribution, MultiRFM excels in estimating study-specified factors and loading matrices compared to other methods. It is also noteworthy that, although RTS demonstrates robustness comparable to MultiRFM, it fails to provide effective information related to the study-specific component.


\begin{table}[H]
    \centering\small
    \caption{
Comparison of MultiRFM and other competing methods for different error distributions in Scenario 2. The reported values are average (standard deviation) of (mean) trace statistics. Note that RTS is unable to estimate the study-specified factor ($\H_s$) and loading ($\B_s$) matrices.}\label{tab:diffdistri}
    \begin{tabular}{llcccccc}
    \hline
        Err & Metric & MultiRFM & MSFA & BMSFA & MSFA-CAVI & MSFA-SVI & RTS \\ \hline
        Gauss & $\mathrm{Tr}_{\A}$ & 0.998 & 0.908 & 0.903 & 0.908 & 0.908 & 0.997 \\   
        ~ & ~ & (2.1e-4) & (2.9e-3) & (3.8e-2) & (2.9e-3) & (3.0e-3) & (8.4e-4) \\   
        ~ & $\mathrm{MTr}_{\F}$ & 0.985 & 0.979 & 0.976 & 0.984 & 0.984 & 0.984 \\   
        ~ & ~ & (3.5e-3) & (1.5e-3) & (2.3e-2) & (3.3e-3) & (3.3e-3) & (3.4e-3) \\   
        ~ & $\mathrm{MTr}_{\B}$ & 0.958 & 0.841 & 0.877 & 0.881 & 0.874 & - \\   
        ~ & ~ & (3.1e-3) & (2.6e-2) & (2.6e-2) & (6.8e-3) & (7.7e-3) & (-) \\   
        ~ & $\mathrm{MTr}_{\H}$ & 0.925 & 0.833 & 0.827 & 0.923 & 0.920 & - \\   
        ~ & ~ & (7.5e-3) & (1.7e-2) & (5.0e-2) & (7.8e-3) & (8.2e-3) & (-) \\   
        \hline
        Exp & $\mathrm{Tr}_{\A}$ & 0.998 & 0.907 & 0.905 & 0.908 & 0.907 & 0.997 \\   
        ~ & ~ & (2.1e-4) & (3.1e-3) & (2.8e-2) & (3.1e-3) & (3.1e-3) & (7.4e-4) \\   
        ~ & $\mathrm{MTr}_{\F}$ & 0.985 & 0.979 & 0.976 & 0.984 & 0.985 & 0.985 \\   
        ~ & ~ & (3.2e-3) & (1.5e-3) & (2.7e-2) & (3.1e-3) & (3.2e-3) & (3.2e-3) \\   
        ~ & $\mathrm{MTr}_{\B}$ & 0.959 & 0.838 & 0.877 & 0.881 & 0.878 & - \\   
        ~ & ~ & (3.2e-3) & (2.6e-2) & (2.0e-2) & (6.1e-3) & (6.3e-3) & (-) \\   
        ~ & $\mathrm{MTr}_{\H}$ & 0.928 & 0.833 & 0.83 & 0.925 & 0.924 & - \\   
        ~ & ~ & (6.7e-3) & (1.8e-2) & (4.0e-2) & (7.1e-3) & (7.1e-3) & (-) \\   
        \hline
        Pareto & $\mathrm{Tr}_{\A}$ & 0.995 & 0.897 & 0.860 & 0.947 & 0.947 & 0.999 \\   
        ~ & ~ & (2.5e-2) & (1.0e-1) & (1.6e-1) & (4.7e-3) & (4.8e-3) & (2.1e-4) \\   
        ~ & $\mathrm{MTr}_{\F}$ & 0.948 & 0.910 & 0.874 & 0.961 & 0.961 & 0.918 \\   
        ~ & ~ & (4.7e-2) & (5.3e-2) & (1.0e-1) & (2.5e-3) & (2.6e-3) & (2.9e-2) \\   
        ~ & $\mathrm{MTr}_{\B}$ & 0.958 & 0.758 & 0.803 & 0.874 & 0.874 & - \\   
        ~ & ~ & (5.6e-2) & (5.0e-2) & (6.6e-2) & (2.3e-2) & (2.4e-2) & (-) \\   
        ~ & $\mathrm{MTr}_{\H}$ & 0.786 & 0.563 & 0.579 & 0.743 & 0.738 & - \\   
        ~ & ~ & (6.4e-2) & (9.4e-2) & (1.1e-1) & (3.8e-2) & (4.1e-2) & (-) \\   
        \hline
    \end{tabular}
\end{table}

\noindent\underline{Scenario 3}. We investigate the influence of the balance in signal strength between study-shared and study-specific factors by examining various combinations of $(\rho_A, \rho_B)$.  We consider three cases, i.e., $(\rho_A, \rho_B)=(2,3), (3,3)$ or $(3, 5)$, with the error distributed according to a multivariate t-distribution with $\nu=3$, while maintaining consistency with the other parameters as outlined in Scenario 1.

Table \ref{tab:signal} illustrates that when the signal strength of the study-specified factor part is held constant at $\rho_B=3$ and the signal strength of the study-shared factor part is increased from $\rho_A=2$ to $\rho_A=3$, there is a noticeable improvement in the estimation accuracy of both the study-shared factor and loading matrices, as well as the study-specified factor and loading matrices across all methods. This enhancement is expected, as an increase in the signal strength of the study-shared factor part leads to more accurate estimates of the study-shared factor and loading matrices. Furthermore, a more precise estimation of the study-shared factor part positively influences the estimation of the study-specific factor part. Conversely, when the signal strength of the study-shared factor part is fixed at $\rho_A=3$ and the signal strength of the study-specific factor part is increased from $\rho_B=3$ to $\rho_B=5$, we observe an improvement in the estimation accuracy of the study-specific factor and loading matrices, but minimal changes in the estimation accuracy of the study-shared factor and loading matrices for nearly all methods. This can be attributed to the fact that an increase in the signal strength of the study-specific factor part does not provide sufficient information to significantly enhance the estimation accuracy of the study-shared factor part.

\begin{table}[H]  
    \centering\small 
    \caption{Comparison of MultiRFM and other competing methods for different signal strengths in Scenario 3. The reported values include average(standard deviation) of  (mean) trace statistics. Note that RTS is unable to estimate the study-specified factor ($\H_s$) and loading ($\B_s$) matrices.}\label{tab:signal} %
   \begin{tabular}{l l cccccc} 
        \toprule  
        \multicolumn{1}{c}{$(\rho_A, \rho_B)$} & {Metric} & {MultiRFM} & {MSFA} & {BMSFA} & {MSFA-CAVI} & {MSFA-SVI} & {RTS} \\  
        \cmidrule(lr){1-8}  
        \multicolumn{1}{c}{(2,3)} & $\mathrm{Tr}_{\A}$ & 0.993 & 0.950 & 0.952 & 0.924 & 0.926 & 0.995 \\  
        & & {(7.6e-3)} & {(4.3e-3)} & {(3.9e-3)} & {(5.8e-2)} & {(5.7e-2)} & {(5.4e-3)} \\  
        & $\mathrm{MTr}_{\F}$ & 0.938 & 0.925 & 0.925 & 0.892 & 0.896 & 0.95 \\  
        & & {(3.3e-2)} & {(1.3e-2)} & {(1.3e-2)} & {(9.8e-2)} & {(9.6e-2)} & {(1.5e-2)} \\  
        & $\mathrm{MTr}_{\B}$ & 0.884 & 0.536 & 0.546 & 0.529 & 0.564 & {-} \\  
        & & {(1.1e-1)} & {(1.5e-1)} & {(1.6e-1)} & {(1.6e-1)} & {(1.5e-1)} & {(-)} \\  
        & $\mathrm{MTr}_{\H}$ & 0.780 & 0.351 & 0.349 & 0.359 & 0.381 & {-} \\  
        & & {(9.9e-2)} & {(1.1e-1)} & {(1.1e-1)} & {(1.2e-1)} & {(1.1e-1)} & {(-)} \\  
        \midrule  
        \multicolumn{1}{c}{(3,3)} & $\mathrm{Tr}_{\A}$ & 0.998 & 0.934 & 0.932 & 0.930 & 0.930 & 0.997 \\  
        & & {(2.4e-4)} & {(4.3e-3)} & {(4.0e-2)} & {(2.5e-2)} & {(2.4e-2)} & {(6.7e-4)} \\  
        & $\mathrm{MTr}_{\F}$ & 0.973 & 0.964 & 0.963 & 0.961 & 0.961 & 0.980 \\  
        & & {(4.2e-3)} & {(6.2e-3)} & {(1.9e-2)} & {(3.9e-2)} & {(3.8e-2)} & {(2.0e-2)} \\  
        & $\mathrm{MTr}_{\B}$ & 0.957 & 0.640 & 0.658 & 0.653 & 0.672 & {-} \\  
        & & {(3.6e-3)} & {(1.2e-1)} & {(1.2e-1)} & {(1.4e-1)} & {(1.2e-1)} & {(-)} \\  
        & $\mathrm{MTr}_{\H}$ & 0.872 & 0.448 & 0.446 & 0.459 & 0.479 & {-} \\  
        & & {(1.3e-2)} & {(1.0e-1)} & {(1.1e-1)} & {(1.2e-1)} & {(1.1e-1)} & {(-)} \\  
        \midrule  
        \multicolumn{1}{c}{(3,5)} & $\mathrm{Tr}_{\A}$ & 0.998 & 0.936 & 0.938 & 0.932 & 0.932 & 0.988 \\  
        & & {(3.0e-4)} & {(3.8e-3)} & {(3.8e-3)} & {(2.5e-2)} & {(2.3e-2)} & {(7.1e-3)} \\  
        & $\mathrm{MTr}_{\F}$ & 0.969 & 0.962 & 0.962 & 0.953 & 0.955 & 0.960 \\  
        & & {(5.1e-3)} & {(6.8e-3)} & {(6.9e-3)} & {(3.9e-2)} & {(3.8e-2)} & {(1.1e-2)} \\  
        & $\mathrm{MTr}_{\B}$ & 0.953 & 0.728 & 0.746 & 0.721 & 0.740 & {-} \\  
        & & {(1.3e-2)} & {(9.4e-2)} & {(9.8e-2)} & {(1.1e-1)} & {(1.0e-1)} & {(-)} \\  
        & $\mathrm{MTr}_{\H}$ & 0.906 & 0.574 & 0.574 & 0.588 & 0.614 & {-} \\  
        & & {(1.1e-2)} & {(1.2e-1)} & {(1.2e-1)} & {(1.3e-1)} & {(1.3e-1)} & {(-)} \\  
        \bottomrule  
    \end{tabular}  
    \label{tab:data} 
\end{table}

\noindent\underline{Scenario 4}. In this Scenario, we examine the performance of the proposed SSVR criterion in Section \ref{sec:select}, in estimating the number of factors. {We maintain consistency with Scenario 1 in terms of parameters except for $(\rho_A,\rho_B)=(6,6)$}. We compare our proposed SSVR criterion with the Akaike and Bayesian
information criteria for MSFA in \cite{de2019multi}, denoted by MSFA-AIC and MSFA-BIC, and eigenvalue ratio based method in RTS~\citep{he2022large}, denoted by RTS-ER. It is noted that RTS-ER is only able to determine the number of study-shared factors.
\begin{figure}
\centering
\includegraphics[width=1\textwidth]{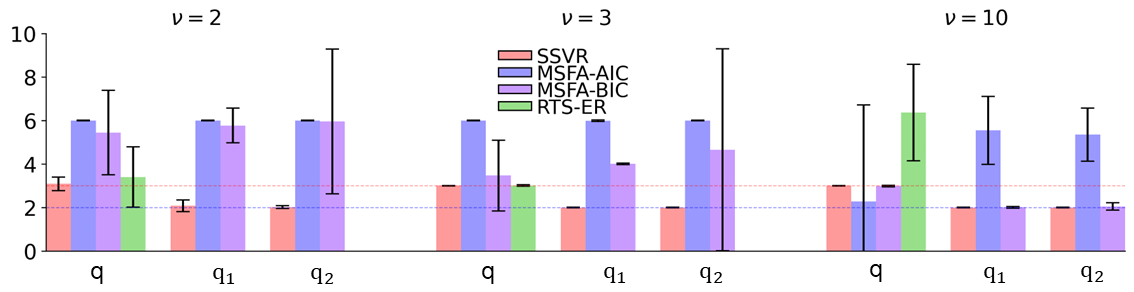}
  \caption{Bar plots of the average of selected $q$ and $(q_1,q_2)$ for the proposed SSVR method and compared methods over 100 repeats, the error bars represent the mean value $\pm$ SD. True values are $q=3$ (red dashed line) and $(q_1,q_2)=(2,2)$ ( blue dashed line).}\label{fig:selectq}
\end{figure}

Figure \ref{fig:selectq} indicates that the SSVR criterion proposed in Section \ref{sec:select} is capable of accurately determining the underlying number of factors across all the cases considered. Furthermore, as the tail of the error becomes lighter, the accuracy of the SSVR in estimating the number of factors increases. In contrast, the AIC and BIC criteria of the MSFA model tend to overestimate both the study-shared and study-specific factors. Although the eigenvalue ratio-based method in RTS performs well in estimating the number of study-shared factors when the error term has heavy tails (i.e., $\nu=2$ or $3$), it fails to account for study-specific factors and tends to overestimate the number of study-shared factors when the error term has lighter tails (i.e., $\nu=20$).

{\noindent\underline{Scenario 5}}. 
Finally, we investigate the influence of variable dimension and the sample size on the parameter estimation of MultiRFM and compared methods. To save space, the simulation settings and results (Table S1--S2) are detailed in Appendix C of Supplementary Materials.
\nvs
\section{Real data analysis}\label{sec:real}
In this section, we present the successful application of MultiRFM in analyzing RNA sequencing data derived from diverse and heterogeneous sources within the biology field. 


\subsection{scRNA-seq data of PBMCs from  COVID-19 patients}
Initially, we utilize the proposed MultiRFM to analyze high-dimensional, multi-study single-cell RNA sequencing datasets of peripheral blood mononuclear cells (PBMCs) from three female subjects: two COVID-19 patients and one healthy individual ~\citep{arunachalam2020systems}. Specifically, the first COVID-19 patient exhibited severe symptoms for 15 days, whereas the second patient displayed moderate symptoms for only two days. After data quality control and {\red log-normalization~[ref:cell]},  we obtain the refined datasets with 4885, 6229, and 5991 cells, respectively. Subsequently, we identify and select the top 2000 highly variable genes common to all three data~\citep{liu2023probabilistic}. Thus, in this data analysis, $S=3, (n_1,n_2, n_3)=(4885, 6229, 5991)$ and $p=2000$.

Before fitting MultiRFM, we use the SSVR  criterion proposed in Section \ref{sec:select} to determine the numbers of study-shared and study-specified factors, resulting $(\hat q, \hat q_1, \hat q_2, \hat q_3)=(6,3,5,4)$. To demonstrate the effectiveness of MultiRFM, we also fit the compared methods using the same number of factors for the fairness of comparison. First, we compare the running time of MultiRFM and other competing methods in Table \ref{tab:real1metric}. For this high-dimensional large-scale data, MultiRFM shows high computational efficiency  spending only 0.39 hour and significantly exceeds other methods. In contrast,  CAVI, SVI and RTS spend more than 4 hours, and MSFA and BMSFA spend 8.22 and 31.56 hours respectively. 

Since there are no true values of model parameters in real data analysis,  we utilize two methods to assess the model fitting performance. First, we propose a metric to measure the reconstruction error for a fitted model. We define $\wh\x_{si}=\wh\bmu_s + \wh\A \wh\f_{si} + \wh\B \wh\h_{si}$ as the reconstruction of the gene expression vector $\x_{si}$ and the reconstruction error of each gene $j$ in study $s$ as $\mathrm{RE}_{sj}=\sqrt{\frac{1}{n_s}\sum_{i} (\wh x_{sij}-x_{sij})^2}$. Then, we regard the average reconstruction error over all genes for each study $s$, given by $\mathrm{RE}_{s}=\frac{1}{p}\sum_j \mathrm{RE}_{sj}$, as the first metric. Second, we define a out-of-sample prediction error by dividing entire data into training  and testing data. Specifically, we first fit a  model based on training data and obtain the estimated parameters, i.e., $(\wh\A^{(train)}, \wh\bmu_s^{(train)}, \wh\B_s^{(train)}, s=1, \cdots, 3)$. Then we define the factor scores for out-of-sample observations: $\wh\f_{si}^{(test)} = (\wh\A^{(train),\trans}\wh\A^{(train)})^{-1}\wh\A^{{(train)},\trans}(\x_{si}^{(test)}-\wh\bmu_s^{(train)})$ and $\wh\h_{si} = (\wh\B_s^{(train),\trans}\wh\B_s^{(train)})^{-1}\wh\B_s^{(train),\trans}(\x_{si}^{(test)}-\wh\bmu_s^{(train)})$ since we do not use the testing data in obtain the parameter estimation. Then we define $\wh\x_{si}^{(test)}=\wh\bmu_s^{(train)} + \wh\A^{(train)} \wh\f_{si}^{(test)} + \wh\B^{(train)} \wh\h_{si}^{(test)}$ as the out-of-sample prediction for $\x_{si}^{(test)}$. Finally, we calculate the average prediction error over all genes for study $s$  
 by $\mathrm{PE}_s=\frac{1}{p}\sum_j \mathrm{PE}_{sj}$, where $\mathrm{PE}_{sj}=\sqrt{\frac{1}{n_{s,test}}\sum_{i} (\wh x_{sij}^{(test)}-x_{sij}^{(test)})^2}$. Table \ref{tab:real1metric} summarizes the results of reconstruction error and prediction error for each study. We observe that robust methods (MultiRFM and RTS) significantly surpasse the non-robust methods (MSFA, BMSFA, CAVI and SVI) in terms of both reconstruction error and prediction error across all studies. Although MultiRFM only slightly outperforms RTS in RE and PE, MultiRFM is much faster than RTS in computation time and is able to provide the study-specific  information about data.
\begin{table}[H]    \centering\caption{Comparison of MultiRFM and competing methods in terms of computational time (Time in hours), reconstruction error (RE) and prediction error (PE) for each study in scRNA-seq data of PBMCs.  The optimal values are highlighted in bold. The standard deviations of RE and PE are also reported in parentheses. Studies 1-3 correspond to the patient with 15 days, the patient with 2 days, and the healthy individual, respectively.}\label{tab:real1metric}
    \begin{tabular}{lccccccc}
    \hline
        Metric & Study & MultiRFM & MSFA & BMSFA & CAVI & SVI & RTS \\ \hline
        Time (hour) & - & {\bf 0.39} & 8.22 & 31.56 & 6.9 & 5.33 & 4.49 \\ \hline
        RE & 1  & {\bf 0.317} & 0.379 & 0.374 & 0.422 & 0.595 & 0.327 \\ 
        ~ & ~ & (0.199) & (0.221) & (0.22) & (0.167) & (0.32) & (0.198) \\ 
        ~ & 2 & {\bf 0.305} & 0.38 & 0.373 & 0.425 & 0.543 & 0.314 \\
        ~ & ~ & (0.201) & (0.234) & (0.235) & (0.16) & (0.265) & (0.2) \\ 
        ~ & 3 & {\bf 0.296} & 0.393 & 0.378 & 0.406 & 0.500 & 0.305 \\ 
        ~ & ~ & (0.196) & (0.241) & (0.221) & (0.167) & (0.245) & (0.197) \\ \hline
        PE & 1 & {\bf 0.314} & 0.364 & 0.353 & 0.34 & 0.34 & 0.327 \\
        ~ & ~ & (0.001) & (0.001) & (0.003) & (0.001) & (0.001) & (0.001) \\ 
        ~ & 2 & {\bf 0.311} & 0.357 & 0.348 & 0.328 & 0.328 & 0.314 \\ 
        ~ & ~ & (0.001) & (0.001) & (0.005) & (0.001) & (0.001) & (0.001) \\ 
        ~ & 3 & {\bf 0.302} & 0.348 & 0.341 & 0.318 & 0.317 & 0.305 \\ 
        ~ & ~ & (0.001) & (0.002) & (0.008) & (0.001) & (0.001) & (0.001) \\ \hline
    \end{tabular}
\end{table}

\begin{figure}
\centering
\includegraphics[height=16cm]{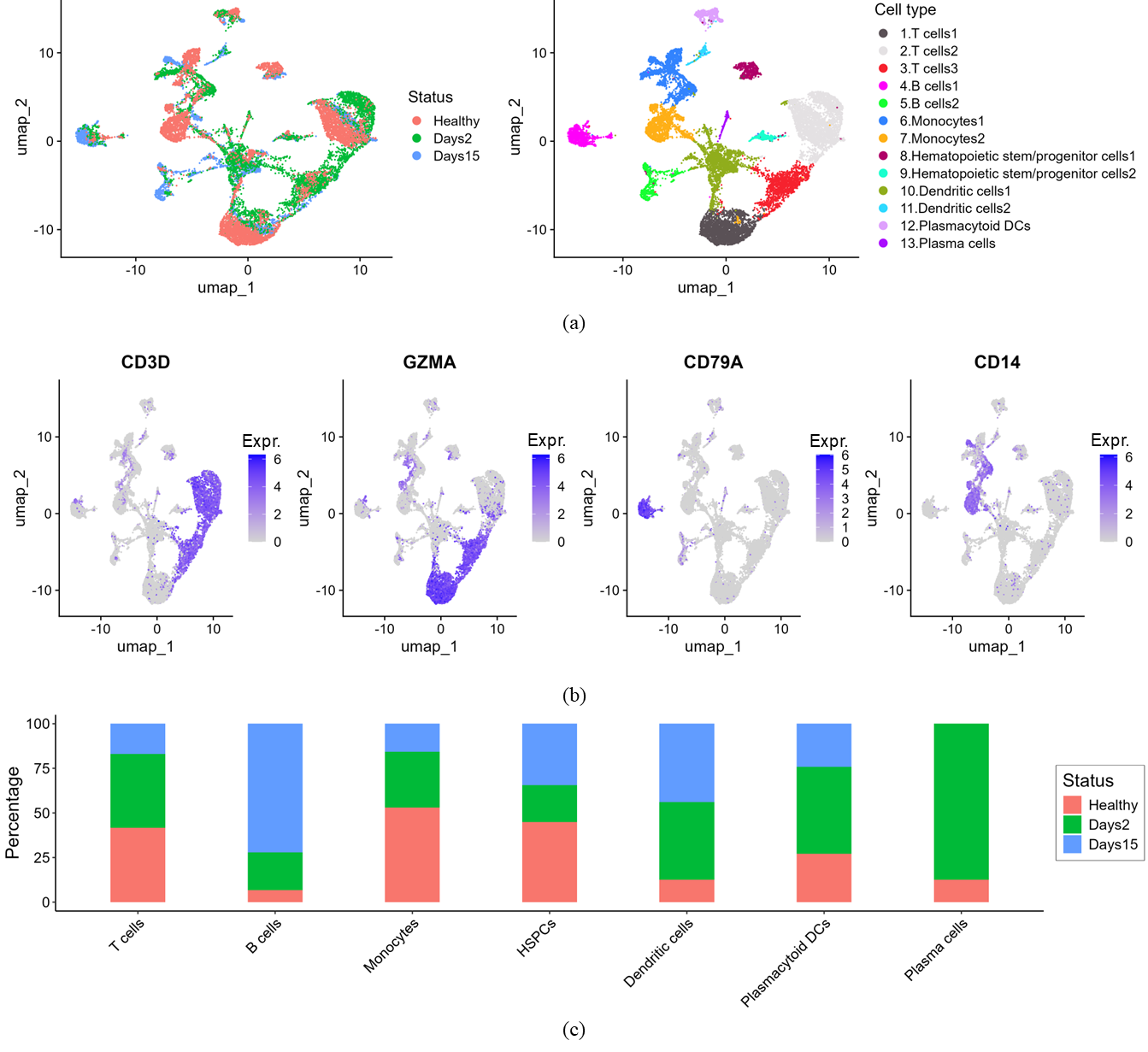}\caption{(a) UMAP plots for visualizing the three individuals and 13 cell clusters, where the UMAP projections are calculated based on the batch-corrected features extracted by MultiRFM.  (b) UMAP plots for visualizing the expression levels of marker genes of T cells ({\it CD3D, GZMA}), Monocytes ({\it CD14}) and B cells ({\it CD79A}). (c) Bar plots of the proportions for each cell type across three individuals, where HSPCs is the abbreviation of hematopoietic stem/progenitor cells.}\label{fig:real1}
\end{figure}

Next, we demonstrate that the features extracted by MultiRFM enhance cell type identification. We define $\wh\V_s = (\F_s, \widetilde\H_s) \in \mathbb{R}^{n\times 11}$, where $\widetilde\H_1 = (\wh\F_1, \wh\H_1, \bbo_{n_1\times 2})$, $\widetilde\H_2 = (\wh\F_2, \wh\H_2)$, and $\widetilde\H_3 = (\wh\F_3, \wh\H_3, \bbo_{n_3\times 1})$. To mitigate batch effects in the features from three individuals, we employ the Harmony method~\citep{korsunsky2019fast} to obtain batch-corrected features. Subsequently, we use Louvain clustering~\citep{blondel2008fast} to jointly cluster cells from three individuals into 13 clusters. In Figure \ref{fig:real1}(a), UMAP plots derived from the batch-corrected features reveal that nearly each cluster is shared among the three individuals, and the clusters are well-separated. To determine the cell type of each cluster, we perform differential gene expression analysis~\citep{stuart2019comprehensive} separately for all clusters from each individual. Using the marker genes of each cluster shared by the three individuals, we identify the cell types for all clusters, as illustrated in Figure \ref{fig:real1}(a) and Supplementary Table S3.  For instance, clusters 1-3 correspond to three distinct subtypes of T cells, identified by the marker genes {\it CD3D, GZMA, CD8A, CD8B}; clusters 4-5 represent two different subtypes of B cells, distinguished by the marker gene {\it CD79A}; and clusters 6-7 are monocytes, recognized by the marker gene {\it CD14}. The marker genes for other clusters are presented in Supplementary Table S3, along with their corresponding cell types. We visualize the gene expression levels of marker genes for each cell type in UMAP plots, as shown in Figure \ref{fig:real1}(b) and Supplementary Figure S1, demonstrating that these marker genes are specifically expressed in their respective cell types. Figure \ref{fig:real1}(c) compares the proportions of various cell types in three individuals. Specifically, the proportion of T cells in the COVID-19 patient with symptoms for 15 days significantly decreases by comparing the healthy and patient with symptoms for two days, indicating the consumption of T cells during the immune response process. Additionally, monocytes, a subset of white blood cells, are also consumed in the immune response process when comparing the three individuals. In contrast, the proportions of B cells show an increasing trend, from the healthy individual to the COVID-19 patient with two days of symptoms, and then to the COVID-19 patient with 15 days of symptoms, suggesting a continuous immune response by generating more B cells.
Finally, we demonstrate that the study-specified loading matrices extracted using MultiRFM enhance gene co-expression network analysis in Appendix D of Supplementary Materials (Figure S2--S3). 

\nvs
\subsection{SRT data of human breast cancer from two patients}
Breast cancer is a significant cause of death among women among various types of cancer. In this application, we use MultiRFM to analyze two spatial resolved transcriptomics (SRT) data of human breast cancers from two patients collected by 10X Visium technology~\citep{he2024starfysh}, corresponding to primary estrogen receptor-positive (ER+) breast cancer and triple-negative
breast cancer (TNBC), respectively. After implementing quality control and selecting highly variable genes for these two datasets, 2454 and 3244 sequencing units (observations) are remained, respectively, both containing 2000 genes for subsequent analysis. 

First, we utilize the SSVR criterion to determine the number of factors, resulting in $(\hat q, \hat q_1, \hat q_2) = (3, 3, 3)$. Then, we compare MultiRFM with other methods. Table \ref{tab:real2metric} indicates that MultiRFM nearly achieves lowest computational time, prediction error, and reconstruction error, with a large margin compared with all non-robust methods (MSFA, BMSFA, CAVI and SVI). Moreover, the robust methods (MultiRFM and RTS) significantly outperforms all non-robust methods in all performance metrics, implying the importance of robustness in analyzing RNA sequencing data.
\begin{table}[H]
    \centering\caption{Comparison of MultiRFM and competing methods in terms of computational time (Time in minutes), reconstruction error (RE) and prediction error (PE) for each study in SRT data of human breast cancer. The optimal values are highlighted in bold. The standard deviations of RE and PE are also reported in parentheses. Studies 1-2 correspond to the ER+ breast cancer and TNBC, respectively.}\label{tab:real2metric}
    \begin{tabular}{lccccccc}
    \hline
        Metric & Study & MultiRFM & MSFA & BMSFA & CAVI & SVI & RTS \\ \hline
        Time (min.) & - & {\bf 17.77} & 123.23 & 613.18 & 26.33 & 109.5 & 45.12 \\ \hline
        RE & 1 & {\bf 0.341} & 0.422 & 0.421 & 0.386 & 0.405 & 0.347 \\ 
        ~ & ~ & (0.229) & (0.282) & (0.287) & (0.215) & (0.235) & (0.234) \\
        ~ & 2 & 0.268 & 0.346 & 0.345 & 0.318 & 0.347 & {\bf 0.254} \\ 
        ~ & ~ & {(0.205)} & (0.234) & (0.235) & (0.175) & (0.199) & (0.185) \\ \hline
        PE & 1 & {\bf 0.341} & 0.383 & 0.371 & 0.356 & 0.357 & 0.349 \\ 
        ~ & ~ & (0.002) & (0.002) & (0.002) & (0.002) & (0.002) & (0.002) \\ 
        ~ & 2 & {\bf 0.249} & 0.283 & 0.272 & 0.262 & 0.262 & 0.256 \\ 
        ~ & ~ & (0.003) & (0.002) & (0.004) & (0.002) & (0.002) & (0.002) \\ \hline
    \end{tabular}
\end{table}

\begin{figure}
\centering
\includegraphics[width=1\textwidth]{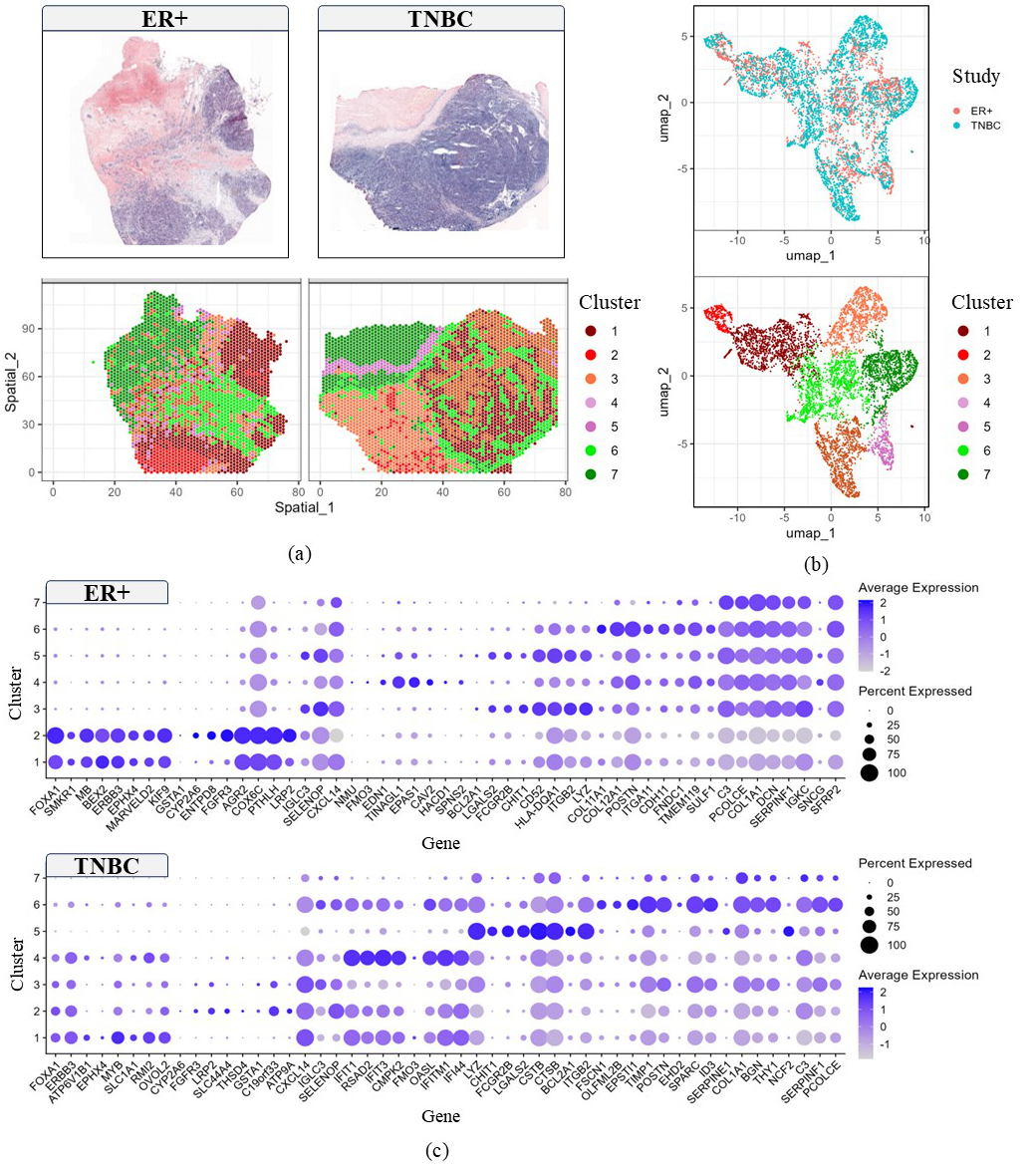}
  \caption{Downstream applications utilizing the features extracted by MultiRFM. (a) upper panel: H\&E staining images of ER+ and TNBC breast cancer tissues; bottom panel: spatial heatmap of the identified clusters based on the batch-corrected features extracted by MultiRFM.  (b) UMAP plot for visualizing the studies and seven cell clusters, where the UMAP projections are calculated based on the batch-corrected features extracted by MultiRFM. (c) Dot plots of the expression levels of top eight marker genes for each cluster in the two tissues.}\label{fig:Real21}
\end{figure}

Next, we demonstrate that the features extracted by MultiRFM, $\wh\V_{s}=(\wh\F_s, \wh\H_s)\in \mathbb{R}^{n_s \times 6}$, facilitate the detection of heterogeneous regions within two breast cancer tissues. 
Using the batch-corrected features, we then perform Louvain clustering to segment the tissues into distinct cellular regions. Figure \ref{fig:Real21}(a) presents H\&E-stained images of two tissue samples in the upper panel, alongside seven identified heterogeneous regions in the lower panel. By comparing these clusters with histology images, we find that clusters 1–5 correspond to five subgroups of cancer cells, while clusters 6 and 7 represent two subgroups of non-cancerous cells. 
We further visualize the batch-corrected features using two-dimensional UMAP projections in Figure \ref{fig:Real21}(b), which illustrate both the segregation of identified cell clusters and the effective integration of data across studies. Interestingly, in the TNBC tissue, cancer cells, corresponding cluster 5, is located between the two parts of the non-cancerous subgroup 7. To better understand the genetic differences between clusters, we perform differential gene expression analysis among clusters in each study. The dot plots in Figure \ref{fig:Real21}(c) display the expression levels of top marker genes for each cluster. Notably, clusters 1 and 2 exhibit similar marker gene expression patterns, suggesting they belong to the same subtype of cancer cells. In both ER+ and TNBC tissues, the genes {\it FOXA1} and {\it ERBB3} emerge as key markers for clusters 1 and 2. 
Consistent with previous studies~\citep{bernardo2013foxa1}, {\it FOXA1} is often highly expressed in ER+ breast cancer, where its expression is closely linked to estrogen receptor status. Conversely, {\it ERBB3}, a member of the HER2 family, may play a role in certain subtypes of TNBC and influence prognosis. Cluster 5 is enriched with genes such as {\it LYZ, LGALS2, CSTB} and {\it ITGB2}. Notably, {\it LYZ} is often overexpressed in tumor-associated macrophages, which can influence tumor progression~\citep{gu2023aberrant}. Meanwhile, {\it ITGB2} plays a crucial role in cell adhesion and migration, processes that are essential for cancer invasion and metastasis~\citep{liu2018lncrna}.

\begin{figure}
\centering
\includegraphics[width=0.8\textwidth]{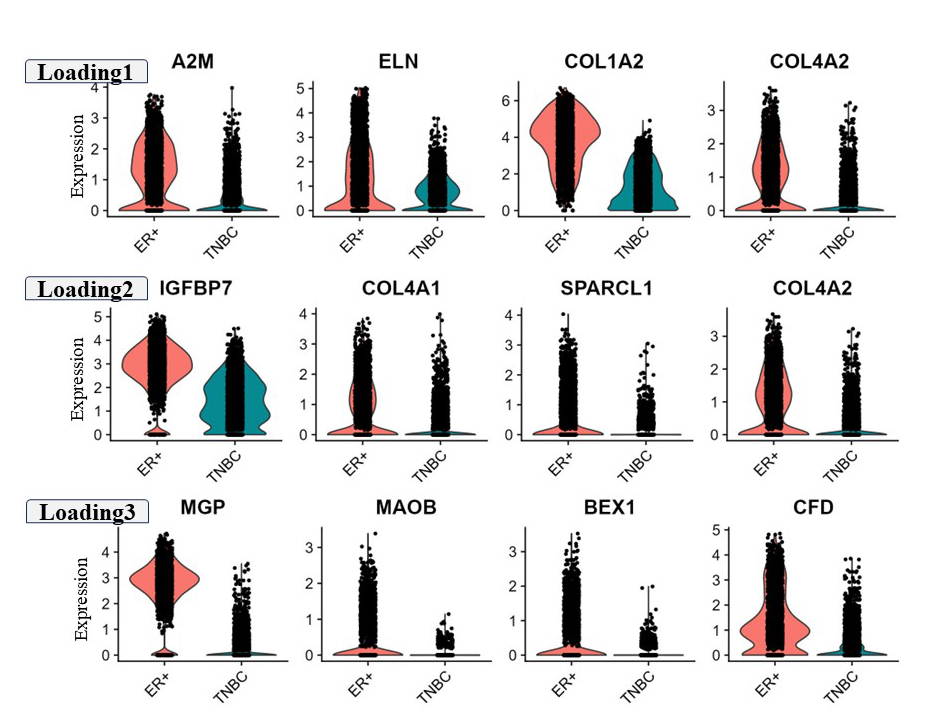}
  \caption{Identification of key genes utilizing the study-specific loadings obtained by MultiRFM. Violin plots illustrating the different expression levels of  four key genes among two types of breast cancers using ER+ breast cancer study-specific loading matrix.}\label{fig:Real22}
\end{figure}

Finally, we demonstrate that the study-specific loading matrices derived from MultiRFM benefit the identification of crucial genes that exhibit differential expression between two studies. We achieve this by ranking the magnitudes of the study-specific loading matrix $\wh\B_s \in \mathbb{R}^{2000\times 3}$. This methodology is driven by the observation that the magnitudes of the loading $\hat b_{sjk}$ indicate the contribution of gene $j$ to the $k$-th study-specific factor. For each of three directions ($q_s=3$), we select the top four genes and visualize their expression levels in both ER+ and TNBC breast cancers. Figure \ref{fig:Real22} demonstrates that the genes specific to the loadings of ER+ breast cancer, including {\it A2M, COL1A2}, {\it COL4A1}, and {\it SPARCL1}, exhibit significantly higher expression levels in ER+ breast cancer compared to TNBC. It is noteworthy that an elevated expression of {\it A2M} has been reported to enhance anti-tumorigenic activity and holds considerable potential as a novel therapeutic agent~\citep{kurz2017anti}. The {\it COL4A1} gene stimulates the growth and metastasis of cancer cells~\citep{wang2020col4a1}. Additionally, the {\it SPARCL1} gene is closely related to the progression, metastasis, and prognosis of various tumor types~\citep{cao2013clinicopathological}. Supplementary Figure S4 showcases genes specific to the loadings of TNBC  that demonstrate greater expression in TNBC than in ER+ breast cancer. These discoveries may offer potential target genes for drug development in both types of breast cancer, providing valuable insights for practitioners.

\section{Discussion}\label{sec:dis}
We have introduced a novel statistical model, named MultiRFM, specifically designed for the robust analysis of noisy RNA sequencing datasets originating from heterogeneous sources. This model is particularly advantageous in scenarios where both the sample size and variable dimension are substantial, and significant technical noise or errors may arise during data collection. 
Under certain regularity conditions, we have proved that the model is identifiable, ensuring the interpretability of its parameters. To address the challenging high-dimensional integral in the observed log-likelihood, we innovatively construct a tractable lower bound function using variational approximation and convex scaling. Leveraging this variational lower bound function, we develop a computationally efficient variational EM algorithm. Furthermore, we employ a step-wise singular value ratio-based criterion, utilizing the robust estimator of loading matrices from MultiRFM, to determine the number of both study-shared and specified factors. Extensive simulation studies reveal that MultiRFM is robust to various error distributions and offers higher estimation accuracy for model parameters compared to existing methods. Two real-data applications further confirm the superior model fitting ability, prediction  capability and computational efficiency of MultiRFM in analyzing noisy, high-dimensional, large-scale RNA sequencing data from multiple heterogeneous sources.

There are three potential extensions to MultiRFM. Firstly, in the presence of additional covariates for sequencing units, we can incorporate the augmented-covariates factor model framework, as outlined in~\cite{liu2024high}, into our robust modeling approach. Secondly, the robustness strategy employed in MultiRFM can be extended to a nonlinear generalized factor model that accommodates a mix of different variable types. Thirdly, exploring how to manage multi-study matrix-variate data from multiple studies using the robust factor model framework presented in this paper is also a promising direction for future research.


\nvs
\section*{Funding}
Liu's work was supported by National Natural Science Foundation of China  (12401361) and  the Fundamental Research Funds for the Central Universities (1082204112J06).
\nvs
\section*{Data availability}
The scRNA-seq data of PBMCs  can be accessed at \url{https://www.ncbi.nlm.nih.gov/geo/query/acc.cgi?acc=GSE155673}. Additionally, the SRT dataset of human breast cancer is available at \url{https://www.ncbi.nlm.nih.gov/geo/query/acc.cgi?acc=GSE218951}.

\nvs
\bibliographystyle{biom}

\bibliography{reflib1}

\begin{thebibliography}{}

\bibitem[\protect\citeauthoryear{Argelaguet, Arnol, Bredikhin, Deloro, Velten, Marioni, and Stegle}{Argelaguet et~al.}{2020}]{argelaguet2020mofa+}
Argelaguet, R., Arnol, D., Bredikhin, D., Deloro, Y., Velten, B., Marioni, J.~C., and Stegle, O. (2020).
\newblock Mofa+: a statistical framework for comprehensive integration of multi-modal single-cell data.
\newblock {\em Genome biology} {\bf 21,} 1--17.

\bibitem[\protect\citeauthoryear{Arunachalam, Wimmers, Mok, Perera, Scott, Hagan, Sigal, Feng, Bristow, Tak-Yin~Tsang, et~al\mbox{.}}{Arunachalam et~al.}{2020}]{arunachalam2020systems}
Arunachalam, P.~S., Wimmers, F., Mok, C. K.~P., Perera, R.~A., Scott, M., Hagan, T., Sigal, N., Feng, Y., Bristow, L., Tak-Yin~Tsang, O., et~al. (2020).
\newblock Systems biological assessment of immunity to mild versus severe covid-19 infection in humans.
\newblock {\em Science} {\bf 369,} 1210--1220.

\bibitem[\protect\citeauthoryear{Bai and Ng}{Bai and Ng}{2013}]{bai2013principal}
Bai, J. and Ng, S. (2013).
\newblock Principal components estimation and identification of static factors.
\newblock {\em Journal of Econometrics} {\bf 176,} 18--29.

\bibitem[\protect\citeauthoryear{Ben-Chetrit, Niu, Swett, Sotelo, Jiao, Stewart, Potenski, Mielinis, Roelli, Stoeckius, et~al\mbox{.}}{Ben-Chetrit et~al.}{2023}]{ben2023integration}
Ben-Chetrit, N., Niu, X., Swett, A.~D., Sotelo, J., Jiao, M.~S., Stewart, C.~M., Potenski, C., Mielinis, P., Roelli, P., Stoeckius, M., et~al. (2023).
\newblock Integration of whole transcriptome spatial profiling with protein markers.
\newblock {\em Nature Biotechnology} {\bf 41,} 788--793.

\bibitem[\protect\citeauthoryear{Bernardo, Bebek, Ginther, Sizemore, Lozada, Miedler, Anderson, Godwin, Abdul-Karim, Slamon, et~al\mbox{.}}{Bernardo et~al.}{2013}]{bernardo2013foxa1}
Bernardo, G.~M., Bebek, G., Ginther, C.~L., Sizemore, S.~T., Lozada, K.~L., Miedler, J.~D., Anderson, L.~A., Godwin, A.~K., Abdul-Karim, F.~W., Slamon, D.~J., et~al. (2013).
\newblock Foxa1 represses the molecular phenotype of basal breast cancer cells.
\newblock {\em Oncogene} {\bf 32,} 554--563.

\bibitem[\protect\citeauthoryear{Blei, Kucukelbir, and McAuliffe}{Blei et~al.}{2017}]{blei2017variational}
Blei, D.~M., Kucukelbir, A., and McAuliffe, J.~D. (2017).
\newblock Variational inference: A review for statisticians.
\newblock {\em Journal of the American statistical Association} {\bf 112,} 859--877.

\bibitem[\protect\citeauthoryear{Blondel, Guillaume, Lambiotte, and Lefebvre}{Blondel et~al.}{2008}]{blondel2008fast}
Blondel, V.~D., Guillaume, J.-L., Lambiotte, R., and Lefebvre, E. (2008).
\newblock Fast unfolding of communities in large networks.
\newblock {\em Journal of statistical mechanics: theory and experiment} {\bf 2008,} P10008.

\bibitem[\protect\citeauthoryear{Cao, Wang, Zhu, Hu, Fang, and Ding}{Cao et~al.}{2013}]{cao2013clinicopathological}
Cao, F., Wang, K., Zhu, R., Hu, Y.-W., Fang, W.-Z., and Ding, H.-Z. (2013).
\newblock Clinicopathological significance of reduced sparcl1 expression in human breast cancer.
\newblock {\em Asian Pacific Journal of Cancer Prevention} {\bf 14,} 195--200.

\bibitem[\protect\citeauthoryear{Cao, Spielmann, Qiu, Huang, Ibrahim, Hill, Zhang, Mundlos, Christiansen, Steemers, et~al\mbox{.}}{Cao et~al.}{2019}]{cao2019single}
Cao, J., Spielmann, M., Qiu, X., Huang, X., Ibrahim, D.~M., Hill, A.~J., Zhang, F., Mundlos, S., Christiansen, L., Steemers, F.~J., et~al. (2019).
\newblock The single-cell transcriptional landscape of mammalian organogenesis.
\newblock {\em Nature} {\bf 566,} 496--502.

\bibitem[\protect\citeauthoryear{Chandra, Dunson, and Xu}{Chandra et~al.}{2024}]{chandra2024inferring}
Chandra, N.~K., Dunson, D.~B., and Xu, J. (2024).
\newblock Inferring covariance structure from multiple data sources via subspace factor analysis.
\newblock {\em Journal of the American Statistical Association} pages 1--15.

\bibitem[\protect\citeauthoryear{De~Vito, Bellio, Trippa, and Parmigiani}{De~Vito et~al.}{2019}]{de2019multi}
De~Vito, R., Bellio, R., Trippa, L., and Parmigiani, G. (2019).
\newblock Multi-study factor analysis.
\newblock {\em Biometrics} {\bf 75,} 337--346.

\bibitem[\protect\citeauthoryear{De~Vito, Bellio, Trippa, and Parmigiani}{De~Vito et~al.}{2021}]{de2021bayesian}
De~Vito, R., Bellio, R., Trippa, L., and Parmigiani, G. (2021).
\newblock Bayesian multistudy factor analysis for high-throughput biological data.
\newblock {\em The annals of applied statistics} {\bf 15,} 1723--1741.

\bibitem[\protect\citeauthoryear{Grabski, De~Vito, Trippa, and Parmigiani}{Grabski et~al.}{2023}]{grabski2023bayesian}
Grabski, I.~N., De~Vito, R., Trippa, L., and Parmigiani, G. (2023).
\newblock Bayesian combinatorial multistudy factor analysis.
\newblock {\em The annals of applied statistics} {\bf 17,} 2212.

\bibitem[\protect\citeauthoryear{Gu, Wang, Dong, Xu, Ye, Shao, Yang, Lu, Chang, Hou, et~al\mbox{.}}{Gu et~al.}{2023}]{gu2023aberrant}
Gu, Z., Wang, L., Dong, Q., Xu, K., Ye, J., Shao, X., Yang, S., Lu, C., Chang, C., Hou, Y., et~al. (2023).
\newblock Aberrant lyz expression in tumor cells serves as the potential biomarker and target for hcc and promotes tumor progression via csgrp78.
\newblock {\em Proceedings of the National Academy of Sciences} {\bf 120,} e2215744120.

\bibitem[\protect\citeauthoryear{Hansen, Avalos-Pacheco, Russo, and De~Vito}{Hansen et~al.}{2024}]{hansen2024fast}
Hansen, B., Avalos-Pacheco, A., Russo, M., and De~Vito, R. (2024).
\newblock Fast variational inference for bayesian factor analysis in single and multi-study settings.
\newblock {\em Journal of Computational and Graphical Statistics} pages 1--42.

\bibitem[\protect\citeauthoryear{He, Jin, Nazaret, Shi, Chen, Rampersaud, Dhillon, Valdez, Friend, Fan, et~al\mbox{.}}{He et~al.}{2024}]{he2024starfysh}
He, S., Jin, Y., Nazaret, A., Shi, L., Chen, X., Rampersaud, S., Dhillon, B.~S., Valdez, I., Friend, L.~E., Fan, J.~L., et~al. (2024).
\newblock Starfysh integrates spatial transcriptomic and histologic data to reveal heterogeneous tumor--immune hubs.
\newblock {\em Nature Biotechnology} pages 1--13.

\bibitem[\protect\citeauthoryear{He, Kong, Yu, and Zhang}{He et~al.}{2022}]{he2022large}
He, Y., Kong, X., Yu, L., and Zhang, X. (2022).
\newblock Large-dimensional factor analysis without moment constraints.
\newblock {\em Journal of Business \& Economic Statistics} {\bf 40,} 302--312.

\bibitem[\protect\citeauthoryear{Jovic, Liang, Zeng, Lin, Xu, and Luo}{Jovic et~al.}{2022}]{jovic2022single}
Jovic, D., Liang, X., Zeng, H., Lin, L., Xu, F., and Luo, Y. (2022).
\newblock Single-cell rna sequencing technologies and applications: A brief overview.
\newblock {\em Clinical and translational medicine} {\bf 12,} e694.

\bibitem[\protect\citeauthoryear{Korsunsky, Millard, Fan, Slowikowski, Zhang, Wei, Baglaenko, Brenner, Loh, and Raychaudhuri}{Korsunsky et~al.}{2019}]{korsunsky2019fast}
Korsunsky, I., Millard, N., Fan, J., Slowikowski, K., Zhang, F., Wei, K., Baglaenko, Y., Brenner, M., Loh, P.-r., and Raychaudhuri, S. (2019).
\newblock Fast, sensitive and accurate integration of single-cell data with harmony.
\newblock {\em Nature methods} {\bf 16,} 1289--1296.

\bibitem[\protect\citeauthoryear{Kurz, Thieme, Amberg, Groth, Jahnke, Pieroh, Horn, Kolb, Huse, Platzer, et~al\mbox{.}}{Kurz et~al.}{2017}]{kurz2017anti}
Kurz, S., Thieme, R., Amberg, R., Groth, M., Jahnke, H.-G., Pieroh, P., Horn, L.-C., Kolb, M., Huse, K., Platzer, M., et~al. (2017).
\newblock The anti-tumorigenic activity of a2m—a lesson from the naked mole-rat.
\newblock {\em PloS one} {\bf 12,} e0189514.

\bibitem[\protect\citeauthoryear{Lafzi, Moutinho, Picelli, and Heyn}{Lafzi et~al.}{2018}]{lafzi2018tutorial}
Lafzi, A., Moutinho, C., Picelli, S., and Heyn, H. (2018).
\newblock Tutorial: guidelines for the experimental design of single-cell rna sequencing studies.
\newblock {\em Nature protocols} {\bf 13,} 2742--2757.

\bibitem[\protect\citeauthoryear{Li and Li}{Li and Li}{2018}]{li2018accurate}
Li, W.~V. and Li, J.~J. (2018).
\newblock An accurate and robust imputation method scimpute for single-cell rna-seq data.
\newblock {\em Nature communications} {\bf 9,} 997.

\bibitem[\protect\citeauthoryear{Liu, Gou, Xia, Wan, Jiang, Sun, Tang, He, and Zhang}{Liu et~al.}{2018}]{liu2018lncrna}
Liu, M., Gou, L., Xia, J., Wan, Q., Jiang, Y., Sun, S., Tang, M., He, T., and Zhang, Y. (2018).
\newblock Lncrna itgb2-as1 could promote the migration and invasion of breast cancer cells through up-regulating itgb2.
\newblock {\em International journal of molecular sciences} {\bf 19,} 1866.

\bibitem[\protect\citeauthoryear{Liu, Liao, Luo, Yang, Lau, Jiao, and et~al.}{Liu et~al.}{2023}]{liu2023probabilistic}
Liu, W., Liao, X., Luo, Z., Yang, Y., Lau, M.~C., Jiao, Y., and et~al. (2023).
\newblock Probabilistic embedding, clustering, and alignment for integrating spatial transcriptomics data with precast.
\newblock {\em Nature Communications} {\bf 14,} 296.

\bibitem[\protect\citeauthoryear{Liu, Lin, Zheng, and Liu}{Liu et~al.}{2023}]{GFMLiu}
Liu, W., Lin, H., Zheng, S., and Liu, J. (2023).
\newblock Generalized factor model for ultra-high dimensional correlated variables with mixed types.
\newblock {\em Journal of the American Statistical Association} {\bf 118,} 1385--1401.

\bibitem[\protect\citeauthoryear{Liu and Zhong}{Liu and Zhong}{2024}]{liu2024high}
Liu, W. and Zhong, Q. (2024).
\newblock High-dimensional covariate-augmented overdispersed poisson factor model.
\newblock {\em Biometrics} {\bf 80,} ujae031.

\bibitem[\protect\citeauthoryear{Liu, DiStasio, Su, Asashima, Enninful, Qin, and et~al.}{Liu et~al.}{2023}]{liu2023high}
Liu, Y., DiStasio, M., Su, G., Asashima, H., Enninful, A., Qin, X., and et~al. (2023).
\newblock High-plex protein and whole transcriptome co-mapping at cellular resolution with spatial cite-seq.
\newblock {\em Nature Biotechnology} pages 1--5.

\bibitem[\protect\citeauthoryear{Nie, Qin, and Liu}{Nie et~al.}{2024}]{nie2024high}
Nie, J., Qin, Z., and Liu, W. (2024).
\newblock High-dimensional overdispersed generalized factor model with application to single-cell sequencing data analysis.
\newblock {\em Statistics in Medicine} .

\bibitem[\protect\citeauthoryear{Qiu, Li, and Yao}{Qiu et~al.}{2024}]{qiu2024robust}
Qiu, J., Li, Z., and Yao, J. (2024).
\newblock Robust estimation for number of factors in high dimensional factor modeling via spearman correlation matrix.
\newblock {\em Journal of the American Statistical Association} pages 1--25.

\bibitem[\protect\citeauthoryear{Shi, Zhou, Jia, Pan, Bai, and Ge}{Shi et~al.}{2021}]{shi2021bias}
Shi, H., Zhou, Y., Jia, E., Pan, M., Bai, Y., and Ge, Q. (2021).
\newblock Bias in rna-seq library preparation: current challenges and solutions.
\newblock {\em BioMed research international} {\bf 2021,} 6647597.

\bibitem[\protect\citeauthoryear{Stuart, Butler, Hoffman, Hafemeister, Papalexi, Mauck~III, and et~al.}{Stuart et~al.}{2019}]{stuart2019comprehensive}
Stuart, T., Butler, A., Hoffman, P., Hafemeister, C., Papalexi, E., Mauck~III, W.~M., and et~al. (2019).
\newblock Comprehensive integration of single-cell data.
\newblock {\em Cell} {\bf 177,} 1888--1902.

\bibitem[\protect\citeauthoryear{Wang, Jin, Hu, Li, Ruan, Xu, Wei, Dong, Teng, Gu, et~al\mbox{.}}{Wang et~al.}{2020}]{wang2020col4a1}
Wang, T., Jin, H., Hu, J., Li, X., Ruan, H., Xu, H., Wei, L., Dong, W., Teng, F., Gu, J., et~al. (2020).
\newblock Col4a1 promotes the growth and metastasis of hepatocellular carcinoma cells by activating fak-src signaling.
\newblock {\em Journal of Experimental \& Clinical Cancer Research} {\bf 39,} 1--16.

\bibitem[\protect\citeauthoryear{Xu, Qi, Li, Yang, Wang, Wang, Liu, Zhao, Liao, Liu, et~al\mbox{.}}{Xu et~al.}{2020}]{xu2020differential}
Xu, G., Qi, F., Li, H., Yang, Q., Wang, H., Wang, X., Liu, X., Zhao, J., Liao, X., Liu, Y., et~al. (2020).
\newblock The differential immune responses to covid-19 in peripheral and lung revealed by single-cell rna sequencing.
\newblock {\em Cell discovery} {\bf 6,} 73.

\bibitem[\protect\citeauthoryear{Yang and Ling}{Yang and Ling}{2023}]{yang2023robust}
Yang, S. and Ling, N. (2023).
\newblock Robust projected principal component analysis for large-dimensional semiparametric factor modeling.
\newblock {\em Journal of Multivariate Analysis} {\bf 195,} 105155.

\bibitem[\protect\citeauthoryear{Ye, Ye, Ye, Li, Ye, Ji, and Wu}{Ye et~al.}{2020}]{ye2020schinter}
Ye, P., Ye, W., Ye, C., Li, S., Ye, L., Ji, G., and Wu, X. (2020).
\newblock schinter: imputing dropout events for single-cell rna-seq data with limited sample size.
\newblock {\em Bioinformatics} {\bf 36,} 789--797.

\bibitem[\protect\citeauthoryear{Yu, He, and Zhang}{Yu et~al.}{2019}]{yu2019robust}
Yu, L., He, Y., and Zhang, X. (2019).
\newblock Robust factor number specification for large-dimensional elliptical factor model.
\newblock {\em Journal of Multivariate analysis} {\bf 174,} 104543.

\end{thebibliography}

\end{document}


%
%

\title{ Supplementary Materials for ``Multi-Study Robust Factor Model for Analyzing RNA Sequencing Data from Heterogeneous Sources" by}
\date{}
\author{Xiaolu Jiang$^{1}$, Wei Liu$^{1*}$\\
$^1$School of Mathematics, Sichuan University, Chengdu,  China\\
*Corresponding author.   Email: \emph{liuwei8@scu.edu.cn}
}

\maketitle
\baselineskip 18pt
\setcounter{section}{0}
\Appendix

\section{Proof of Proposition 1}
{\bf Proof}. Since $\mathrm{E}(\f_{si})=\mathrm{E}(\h_{si})=\mathrm{E}(\beps_{si})=\bbo$, we take expectation on both sides of model (1) in the main text, obtaining $\bmu_s = \mathrm{E}(\x_{si})$. Thus, $\bmu_s$ is identifiable for $s=1, \cdots, S$. Let $\widetilde\B_s=(\A, \B_s) \in \mathbb{R}^{p\times (q+q_s)}$ and $\widetilde\Lambda_s = \frac{\nu}{\nu-2}\Lambda_s$. Leveraging the independence of $\f_{si}, \h_{si}$ and $\beps_{si}$, and model (1) in the main text, we have $\mathrm{cov}(\x_{si}) = \widetilde\B_s \widetilde\B_s^{\trans} + \widetilde\Lambda_s$. Let $\D_s=\widetilde\B_s \widetilde\B_s^{\trans}$. Assuming two sets of parameters $(\D_s^{(1)}, \widetilde\Lambda_s^{(1)})$ and $(\D_s^{(2)}, \widetilde\Lambda_s^{(2)})$ exists, we have 
$$\D_s^{(1)} + \widetilde\Lambda_s^{(1)}=\D_s^{(2)} + \widetilde\Lambda_s^{(2)}.$$
Furthermore, 
\begin{equation}\label{eq:D12}
    \D_s^{(1)} =\D_s^{(2)} + \widetilde\Lambda_s^{(2)}-  \widetilde\Lambda_s^{(1)}.
\end{equation}
Our first objective is to show $(\D_s^{(1)}, \widetilde\Lambda_s^{(1)})=(\D_s^{(2)}, \widetilde\Lambda_s^{(2)})$. There are two cases that may satisfy Equation \eqref{eq:D12}: (a): $\widetilde\Lambda_s^{(1)}=\widetilde\Lambda_s^{(2)}$;  (b): $\widetilde\Lambda_s^{(1)}\neq \widetilde\Lambda_s^{(2)}$.
 Case (a) is trivial, so we focus on the second  case.

By Condition (A1) in the main text, we have $\mathrm{rank}(\D_s^{(1)})=\mathrm{rank}(\D_s^{(2)})=q+q_s$. If Case (b) holds, we perform singular value decomposition (SVD): $\D_s^{(2)}=\U_s\S_s\U_s^{\trans}$, where $\S_{s}$ is a diagonal matrix with non-zero entries only in the first $q+q_s$ diagonal positions. Without loss of generality, we assume $\tilde\lambda_{s1}^{(1)}\neq \tilde\lambda_{s1}^{(2)}$. Then, we have
\begin{equation}\label{eq:K1}
   \U_s^{\trans}\D_s^{(1)}\U_s = \S_s + \K_s, 
\end{equation}
where $\K_s= (\tilde\lambda_{s1}^{(2)}-\tilde\lambda_{s1}^{(1)}) \u_{s1}\u_{s1}^{\trans}$ is a rank-one matrix and $\u_{s1}$ is the first column of $\U_s$. Thus, we conclude that $\S_s+\K_s$ is a rank $q+q_s+1$ matrix, which contracts that $\mathrm{rank}(\D_s^{(1)})=q+q_s$. Therefore, we conclude $(\D_s^{(1)}, \widetilde\Lambda_s^{(1)})=(\D_s^{(2)}, \widetilde\Lambda_s^{(2)})$.

We now further establish the identifiability of $\widetilde\B_{1}$. Assume that $\widetilde\B_{1}$ is not identifiable, then $\D_{1} = \widetilde\B_{1}^{(1)}\widetilde\B_{1}^{(1), \trans} = \widetilde\B_{1}^{(2)}\widetilde\B_{1}^{(2), \trans}$. Following the same arguments in the proofs of Theorem 1 in \cite{liu2024highdimensional}, we have that $\widetilde\B_{1}$ is  identifiable. Thus, both $\A$ and $\B_{1}$ are identifiable. Applying Condition (A3) in the main text and the same arguments of proving $\B_{1}$ to $\{\B_{s},s=2,\cdots, S\}$, we conclude that $\{\B_{s},s=2,\cdots, S\}$ are identifiable.

Finally, we prove the identifiability of $(\nu, \Lambda_s)$. Since $\widetilde\Lambda_s$ is identifiable, we have $\widetilde\Lambda_s=\frac{\nu}{\nu-2}\Lambda_s $. Thus, $\Lambda_s = \frac{\nu-2}{\nu} \widetilde\Lambda_s$. Assuming there are $\nu^{(1)}$ and $\nu^{(2)}$ satisfying model, we have, for any $\beps$, $C_p(\nu^{(1)}) |\frac{\nu^{(1)}-2}{\nu^{(1)}}\widetilde\Lambda_s|^{-1/2} (1+(\nu^{(1)}-2)\beps^{\trans}\widetilde\Lambda_s^{-1} \beps)^{-(\nu^{(1)}+p)/2}=C_p(\nu^{(2)}) |\frac{\nu^{(2)}-2}{\nu^{(2)}}\widetilde\Lambda_s|^{-1/2} (1+(\nu^{(2)}-2)\beps^{\trans}\widetilde\Lambda_s^{-1} \beps)^{-(\nu^{(2)}+p)/2}$. By the arbitrary of $\beps$, we conclude $\nu^{(1)}=\nu^{(2)}$. That is, $\nu$ is identifiable. Furthermore, $\Lambda_s$ is also identifiable.

\section{Pseudocode of MultiRFM}
\begin{algorithm}
	\caption{Pseudocode for MultiRFM}
	\label{alg:vem}
	\begin{algorithmic}[1]
		\Require $\{\X_s, q_s, s\leq S\}$, $q$, maximum iterations $maxIter$, relative tolerance of the variational lower bound value ($eps$).
		\Ensure  $\wh\bg$ and $\wh\btheta$
		\State Initialize  $\bg^{(0)}$ and $\btheta^{(0)}$.
		\For{ each $t = 1, \cdots, maxIter $}
        \State Replace $\phi_{si}(\btheta, \bg)$ in Equations \eqref{eq:Sf}--\eqref{eq:Lambdas} with $\phi_{si}(\btheta^{(t-1)}, \bg^{(t-1)})$;
		\State Update variational parameters $\bg^{(t)}$ based on Equations  \eqref{eq:Sf}--\eqref{eq:mh};
		\State Update model parameters $\btheta^{(t)}$ based on Equations  \eqref{eq:bmu}--\eqref{eq:tau};
        \State Evaluate the variational lower bound $l_t= l(\btheta^{(t)}, \bg^{(t)})$ by  \eqref{eq:lbtbg}.
        \If {$|l_t - l_{t-1}|/ |l_{t-1}| < eps$}
       \State  break;
        \EndIf
		\EndFor
		\State \textbf{return} $\wh\bg=\bg^{(t)}$ and $\wh\btheta=\btheta^{(t)}$.
	\end{algorithmic}
\end{algorithm}

\section{Additional results in real data analysis}

\begin{table}[H]
\centering\small
\renewcommand\tabcolsep{2.2pt} 
\caption{Identification of cell types for 13 clusters based on distinctive marker genes, where HSPCs and pDCs are the abbreviation of hematopoietic stem/progenitor cells and plasmacytoid DCs, respectively.}
\begin{tabular}{cccccccc}
\hline
Cluster & 1 & 2 & 3 & 4 & 5 & 6 & 7   \\
\hline
Cell type & T cells1 & T cells2 & T cells3 & B cells1 & B cells2 & Monocytes1 & Monocytes2\\ \hline
MG & GNLY & CD3D & CD8A & CD79A & CD79A & CD14 & CD14   \\
\multirow{4}{*}{} & CCL5 & CD3E & CD8B & & &  \\
&  NKG7 & & GZMA& & & &\\
& GZMA & & & & & &\\
\hline
Cluster & 8 & 9 & 10 & 11 & 12 & 13 & \\
\hline 
Cell type & HSPCs1 & HSPCs2 & Dendritic cells1 & Dendritic cells2 & pDCs & Plasma cells &\\ \hline 
MG & GATA2 & GATA2 & ITGAX & CD1E & IL3RA & IGHG1 &\\
\multirow{3}{*}{}  & CD34 & & & CD1C & & IGHG2\\
& & & & & & IGKC &\\
\hline
\end{tabular}

\end{table}
\begin{figure}[H]
\centering
\includegraphics[width=0.8\textwidth]{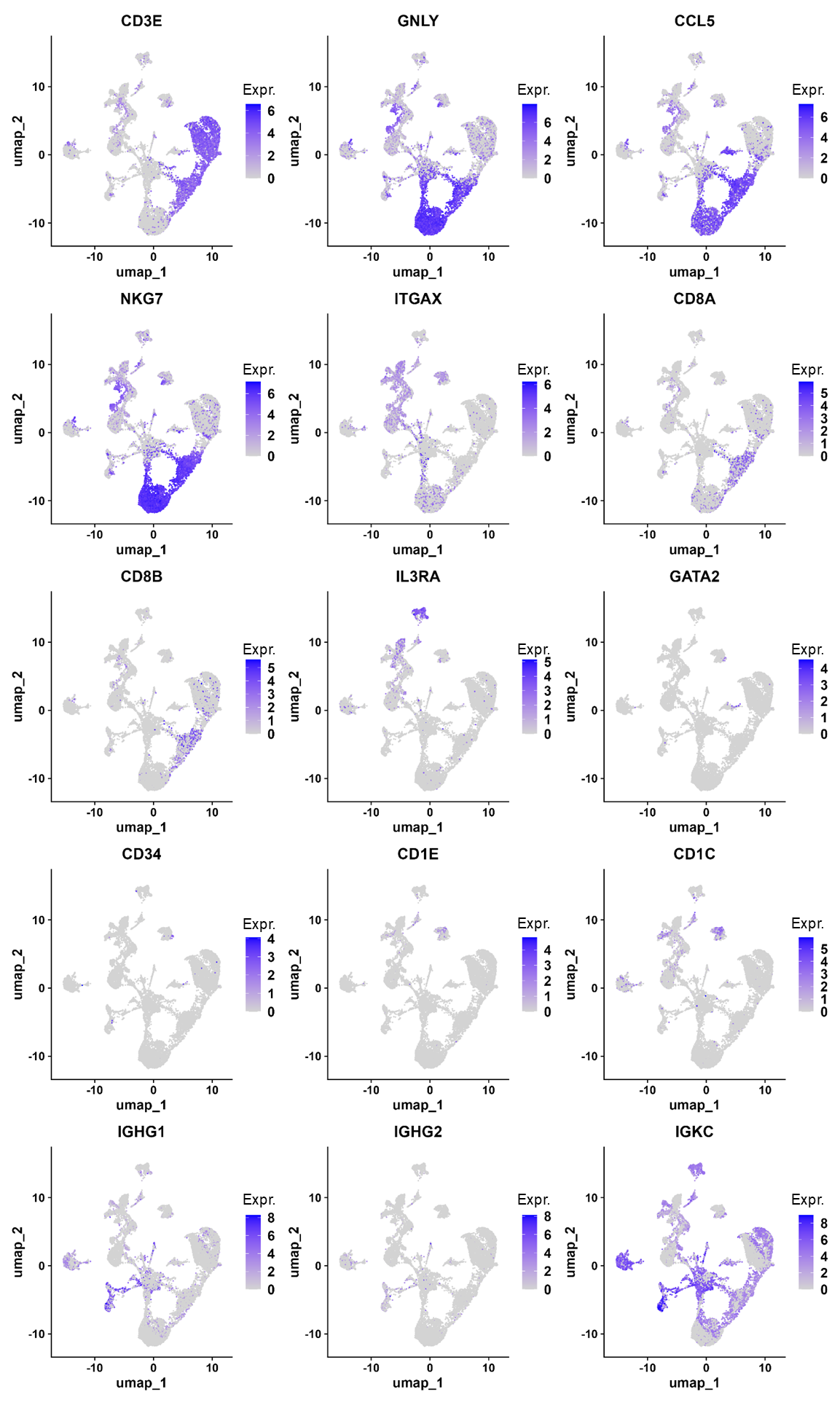}\caption{UMAP plots for visualizing the expression levels of marker genes of identified cell types.}
\end{figure}

\begin{figure}[H]
\centering
\includegraphics[width=1\textwidth]{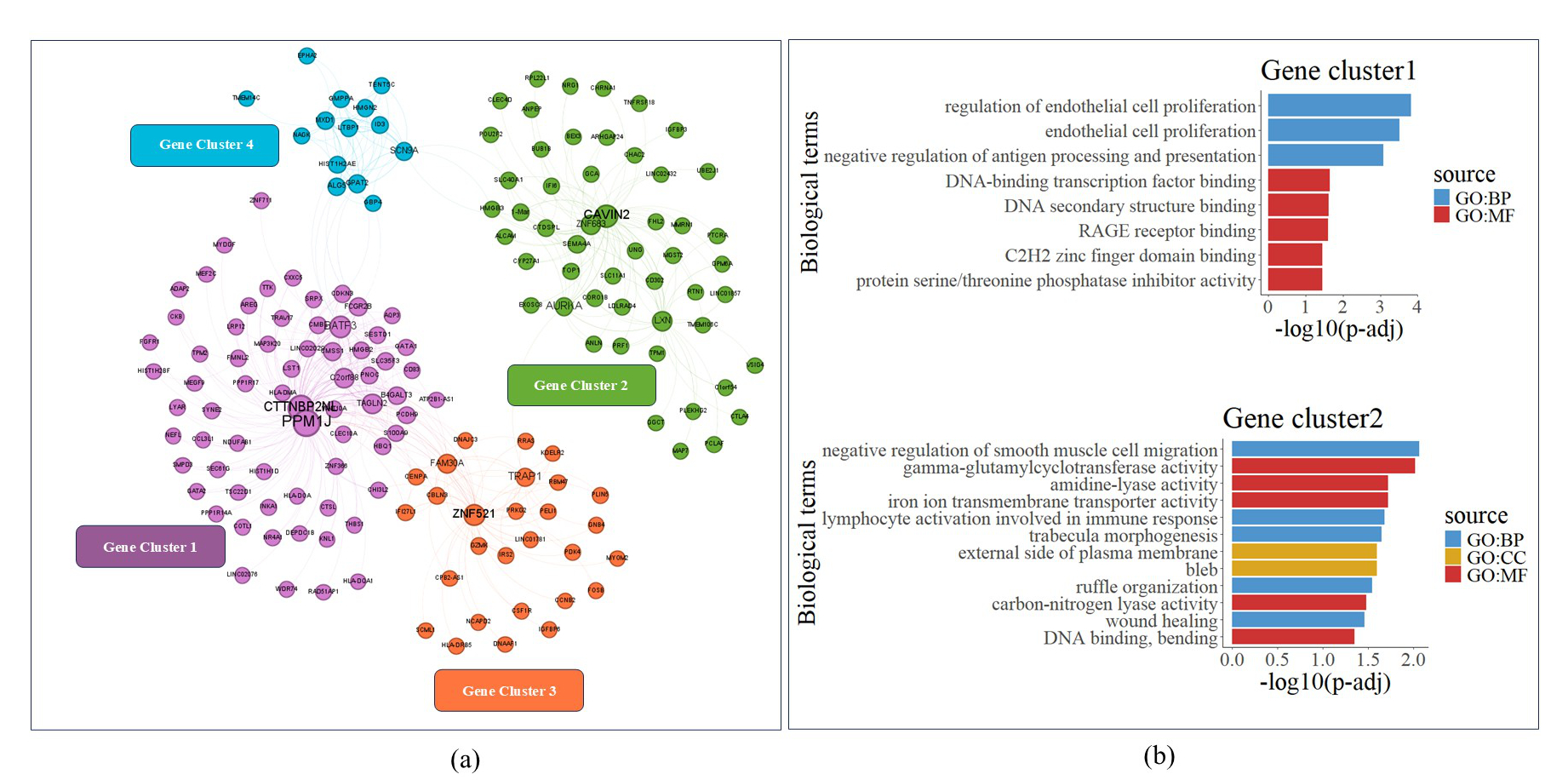}\caption{(a): Gene co-expression networks for the healthy. (b): Top significant pathways in GO database  for the gene clusters 1--2 of the healthy.}\label{fig:realsup2}
\end{figure}

\begin{figure}[H]
\centering
\includegraphics[width=0.8\textwidth]{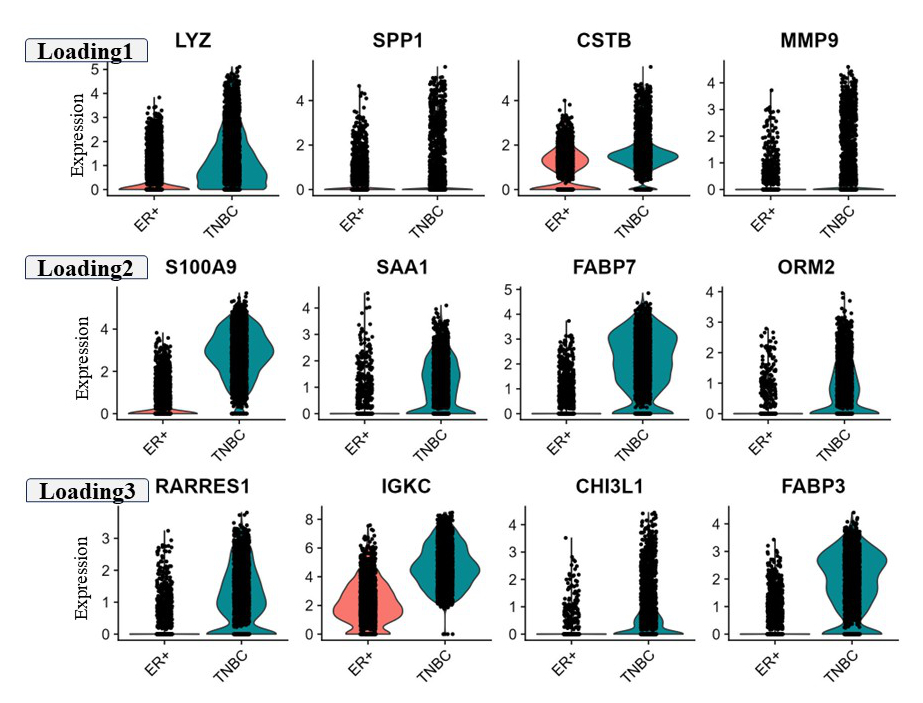}
  \caption{ Identification of key genes utilizing the study-specific loadings obtained by MultiRFM. Violin plots illustrating the different expression levels of four key genes among two
types of breast cancers using TNBC study-specific loading matrix.}\label{fig:Real22}
\end{figure}

\bibliographystyle{apalike}

\bibliography{reflib1}